\documentclass[useAMS,usenatbib]{mn2e}
\usepackage{lscape}
\usepackage{amsmath}
\usepackage{latexsym}
\usepackage{graphicx}
\usepackage{amssymb}

\newcommand{\lsim}{\lower0.6ex\vbox{\hbox{$ \buildrel{\textstyle
        <}\over{\sim}\ $}}}
\newcommand{\gsim}{\lower0.6ex\vbox{\hbox{$ \buildrel{\textstyle
        >}\over{\sim}\ $}}}
\newcommand{\hmpc}{ h^{-1}{\rm Mpc}}\newcommand{\Dvir}{ \Delta_{\rm vir}}

\newcommand{\Vmax}{V_{\rm max}}
\newcommand{\Vin}{V_{\rm in}}

\newcommand{\Nc}{N_{c}}
\newcommand{\hkpc}{ h^{-1}{\rm kpc}}
\newcommand{\Msun}{M_{\odot}}
\newcommand{\hMsun}{h^{-1} \Msun}
\newcommand{\hMpc}{h^{-1} {\rm Mpc}}

\newcommand{\lcdm}{$\Lambda$CDM }
\newcommand{\kms}{km s$^{-1}$}

\newcommand{\Rvir}{R_{\rm vir}}
\newcommand{\beq}{\begin{equation}}
\newcommand{\eeq}{\end{equation}}

\newcommand{\lya}{Ly$\alpha$ }

\newcommand{\kpc}{$h^{-1}$kpc }

\title[Close galaxy pairs at $z=3$]{Close galaxy pairs at $z=3$: A
challenge to UV luminosity abundance matching}

\author[Berrier \& Cooke]{Joel C. Berrier$^{1}$\thanks{E-mail:
jberrier@uark.edu} \& Jeff Cooke$^{2,}$$^{3}$\\ $^{1}$Department of
Physics, University of Arkansas, 835 West Dickson Street,
Fayetteville, AR 72701;\\ and Arkansas Center for Space and Planetary
Sciences, 202 Old Museum Building, University of Arkansas,
Fayetteville, AR 72701\\ $^{2}$California Institute of Technology,
1200 East California Boulevard, Pasadena, CA 91125,
USA\\ $^{3}$Swinburne University of Technology, PO Box 218, Mail Number
H39, Hawthorn, VIC 3122, Australia}

\begin{document}

\date{Accepted 2012 August 6.  Received 2012 July 28; in original form 2011 April 8 \\ The definitive version may be found at:\\ http://onlinelibrary.wiley.com/doi/10.1111/j.1365-2966.2012.21866.x/pdf}
\pagerange{\pageref{firstpage}--\pageref{lastpage}} \pubyear{2012}

\maketitle

\label{firstpage}
 
\begin{abstract} 

We use  a sample  of $z\sim3$ Lyman  Break Galaxies (LBGs)  to examine
close pair clustering statistics in comparison to LCDM-based models of
structure formation.  Samples are  selected by matching the LBG number
density,  $n_g$, and  by  matching the  observed  LBG 3-D  correlation
function  of  LBGs  over  the  two-halo term  region.   We  show  that
UV-luminosity abundance  matching cannot reproduce  the observed data,
but if  subhalos are  chosen to reproduce  the observed  clustering of
LBGs we are able to reproduce the observed LBG pair fraction, ($N_c$),
defined as  the average number  of companions per galaxy.   This model
suggests an over  abundance of LBGs by a factor  of $\sim5$ over those
observed, suggesting  that only  $1$ in $5$  halos above a  fixed mass
hosts  a  galaxy  with  LBG-like  UV  luminosity  detectable  via  LBG
selection  techniques.   This overdensity  is  in  agreement with  the
results of a Millennium 2 analysis and with the discrepancies noted by
previous  authors using  different types  of simulations.   We  find a
total observable close  pair fraction of $23 \pm  0.6$ per cent ($17.7
\pm  0.5$)  per cent  using  a  prototypical  cylinder radius  in  our
overdense fiducial model and $8.3 \pm 0.5$ per cent ($5.6 \pm 0.2$ per
cent)  in an abundance  matched model  (impurity corrected).   For the
matched spectroscopic slit analysis, we find $N_{cs}(R) = 4.3 \pm 1.55
(1.0 \pm 0.2 )$ per cent and  $N_{cs} = 5.1 \pm 0.2$ ($1.68 \pm 0.02$)
per cent, the average number of companions observed serendipitously in
randomly  aligned spectroscopic slits,  for fiducial  slits (abundance
matched), whereas the observed fraction of serendipitous spectroscopic
close pairs  is $4.7 \pm 1.5$ per  cent using the full  LBG sample and
$7.1 \pm  2.3$ per  cent for a  subsample with  higher signal-to-noise
ratio.  We conduct the same analysis on a sample of dark matter haloes
from the Millennium  2 simulation and find similar  results.  From the
results and an analysis of the observed LBG 2-D correlation functions,
we show that the standard method of halo assignment fails to reproduce
the break, or  up turn, in the LBG close pair  behavior at small scale
($\lesssim 20 ~\hkpc$ physical).   To reconcile these discrepancies we
suggest that  a plausible fraction of  LBGs in close  pairs with lower
mass (higher  density) than our  sample experience interaction-induced
enhanced star  formation that boosts their  luminosity sufficiently to
be  detected in  observational  sample  but are  not  included in  the
abundance matched simulation sample.
 
\end{abstract}


%----------------------------------------------------------
\section{INTRODUCTION}
\label{sec:intro}%Section 1
%----------------------------------------------------------

One of the fundamental predictions of a \lcdm model of the universe is
the hierarchical growth of structure.  However, direct observations of
galaxy mergers,  and, by extension, statistics on  galaxy mergers, are
difficult to obtain  due to the long time-scales  of the galaxy-galaxy
merger process.  Correlations between galaxy characteristics and their
environment suggest  that interactions play  a role in  setting galaxy
properties   such   as  star   formation   rate,  colour,   morphology
\citep[e.g.][]{ToomreToomre72,   LarsonTinsley78,   Dressler80,  PG84,
  Barton:00,  Barton:03}.  However,  observational studies  of mergers
and interactions can be difficult due to the low luminosities of tidal
features and the difficulties  in quantifying galaxy morphologies.  At
high  redshifts, $z  \ge 1$,  these  problems are  exacerbated by  the
decreased  apparent luminosity  and resolution  of the  galaxies being
studied.

Since the studies of \citet{Holmberg1937} close pairs of galaxies have
provided an important  tool for the evaluation of  galaxy merger rates
by providing counts  of merger candidates, and for  theories of galaxy
formation  due to the  importance of  galaxy-galaxy mergers  in galaxy
evolution.   Close galaxy  pair counts,  or counts  of morphologically
disturbed systems, have  not only been used to  provide candidates for
galaxy mergers,  but have  been used in  attempts to probe  the galaxy
merger  rate   and  its  evolution   with  redshift  \citep[][]{z&k89,
  Burkey94,  Carlberg94, Woods95,  Y&E95,  P97, Neus97,  Carlberg2000,
  LeFevre2000,  P2002,  Conselice03,  Bundy04, masjedi05,  Belletal06,
  Lotz06, Lin04}.

In \citet{Berrieretal06} we present a method to analyse the close pair
fraction of galaxies in a  simulation environment, with the close pair
fraction ($N_c$) defined as the number of galaxies in close pairs in a
volume  of space normalised  by the  total number  of galaxies  in the
sample.  This  analysis demonstrates  the viability of  estimating the
observable close pair fraction in simulations using simple criteria to
assign galaxies to dark matter haloes.

\citet{Berrieretal06}  argue that the  close luminous  companion count
per galaxy does  not track the distinct dark  matter halo merger rate.
Instead, it  tracks the luminous  galaxy merger rate.  While  a direct
connection between the two has often been assumed, there is a mismatch
because multiple galaxies  may occupy the same host  dark matter halo.
The same  arguments apply  to morphological identifications  of merger
remnants, which also  do not directly probe the  host dark halo merger
rate.   This still leaves  close galaxy  pairs as  a tracer  of galaxy
evolution and as a proxy of the galaxy merger rate.

At high redshift, the dense environment and smaller fraction of galaxy
clusters   [where  the   large  velocity   dispersion   prevents  many
  satellite-satellite mergers, e.g. \citet{Berrieretal2009}] mean that
close pairs of galaxies are  likely to indicate actual mergers, though
estimates of the timescales of these mergers may still be rather large
\citep[see   e.g.,][]{Kitzbichler2008,  Bertone09}.    As   a  result,
observations of this process for  galaxies at high redshift are highly
desirable  for  constraining the  high  redshift galaxy-galaxy  merger
rate.

The Lyman break galaxies  (LBGs) are star forming galaxies efficiently
identified using  colour selection criteria \citep[e.g.,][]{Steidel96}
and  comprise  a large  fraction  of  all  luminous galaxies  at  high
redshift  \citep[e.g.,][]{Reddy05,Marchesini2007}.   To  date,  a  few
thousand LBG  spectra and tens of thousands  of photometric candidates
have been obtained, making  LBGs a useful, well-studied population for
high  redshift galaxy  spatial distribution  and close  pair analysis.
\citet{Conroy2006} used  subhalo abundance matching  techniques (SHAM)
such  as those used  in \citet{Berrieretal06}  and here,  to calculate
angular  correlation functions  for LBGs  at high  redshifts, $z=3,4$.
This work suggests  that abundance matching techniques may  be used to
sample LBG  populations and statistics in simulations.   SHAM has been
tested in  a variety of situations  at both low and  high redshift and
has  been shown  to  be a  reasonable  tool for  matching galaxies  to
populations of  dark matter halos  and generating halo mass  - stellar
mass  relations  \citep{Conroy2006, Vale&Ostriker2006,  Berrieretal06,
  Stewart2009,   Simha2010,   Guo2010,   Moster2010}.   Because   this
technique  has been  suggested, and  indeed used,  as a  probe  of LBG
clustering  statistics, it  will provide  our starting  point  in this
analysis.   We also  explore  matching dark  matter  halo and  subhalo
correlation   functions  to  the   observed  clustering   of  $z\sim3$
LBGs. This technique  is similar to the work  of \citet{Conroy2008} on
$z\sim2$ star forming galaxies.

In  this  paper,  we  use  a numerical  $N-$body  simulation  with  an
analytically   generated  substructure,   adopting  the   approach  of
\citet{Berrieretal06}, to  compare close companion  counts directly to
the observed  companion count for our  sample of $z \sim  3$ LBGs from
the survey  of \citet[][hereafter S03]{Steidel2003} and  the survey of
\citet[][hereafter C05]{Cooke05}.   Our purpose is to  test the simple
and   popular  \citep{Conroy2008,Stewart2009,Simha2010}   theory  that
galaxies  live   in  subhalos   and  that  UV   luminosity  correlates
monotonically with halo mass/maximum  circular velocity at the time of
accretion.

The structure  of this paper is as  follows.  In \S~\ref{sec:methods},
we  outline   our  methods,   discuss  our  observational   sample  in
\S~\ref{sec:observations},          our         simulations         in
\S~\ref{sec:simulations},  the  models  used  for  the  assignment  of
galaxies to haloes in \S~\ref{sec:model}. The definitions of the close
companion fraction, the photometric companion fraction, and the sample
impurity  and   number  density  are  covered   in  \S~\ref{sec:nc}  -
\S~\ref{sub:NG}  respectively.   We present  our  predictions for  the
companion fraction, $\Nc$, in  \S~\ref{sec:results}.  We begin with an
examination of  $N_c$ from  $z=0-3$ with an  emphasis on  a comparison
between  our simulations  and  the observational  values  at $z=3$  in
\S~\ref{sub:evolution}.  The angular photometric close companion count
is  the  topic  of  \S~\ref{sub:pred}.   Comparisons  with  previously
existing close companion counts  are made in \S~\ref{sub:compare}.  We
return to the number  density issue in \S~\ref{sub:Density}.  Finally,
we discuss the implications of our results in \S~\ref{sub:Discussion}.
We conclude with a summary in \S~\ref{sec:conclusions}.

In this  work we assume a  flat universe with a  standard cosmology of
$\Omega_m= 1 - \Omega_{\Lambda} = 0.3$, $h=1.0$, and $\sigma_8 = 0.9$.

%*********************************************************************

%----------------------------------------------------------
\section{METHODS}
\label{sec:methods}%Section 2
%----------------------------------------------------------

Pair  count  statistics are  generated  using  the  same technique  as
\citet{Berrieretal06}.   A  \lcdm   $N$-body  simulation  is  used  to
identify  the large-scale structure  and properties  of the  host dark
matter  halo  (details  in  \S~\ref{sec:simulations}).   The  analytic
substructure model of \citet{Zentner05}  is used to generate four sets
of  satellite galaxies  within  these host  haloes  for our  analysis.
Using the analytic models with  no inherent resolution limits to model
substructure allows us to overcome the issue of numerical over-merging
in the dense environments \citep[e.g.,][]{klypin_etal:99}. This method
has  been demonstrated  to accurately  model the  two-point clustering
statistics  of  haloes and  subhaloes  \citep{Zentner05}  and used  to
produce viable close  pair statistics \citep{Berrieretal06} from $z=0$
to $z=1$.

We use a simple method to assign galaxies to dark matter haloes in our
simulation volume (\S~\ref{sec:model}) and address possible effects of
this   assignment  in   \S~\ref{sub:Discussion}.    We  conduct   mock
observations  on the  ``galaxy'' catalogs  in an  identical  manner as
those used in observational studies to calculate the average number of
close companions, $N_c$,  or the close pair fraction  statistic in our
simulation box (\S~\ref{sec:nc}).  This can be done to mimic the exact
specifications  of observations  in the  real Universe.   The analytic
subhalo  model allows us  to examine  the variance  in close-companion
counts  associated with  the  realisation-to-realisation scatter.   In
this  way we may  examine different  sets of  substructure populations
while retaining  the large-scale structure in  our simulation allowing
us  to  test  for  the   importance  of  cosmic  variance  and  chance
projections.

In  this  work  we focus  on  examining  the  close pair  fraction  of
potential  LBG haloes at  $z=3$ in  a simulation  box and  make direct
comparisons  to  sets  of   observational  data.   In  order  to  more
accurately  test  the expectations  of  detecting serendipitous  close
pairs in conventional multi-object spectroscopic surveys, we calculate
both a  standard $N_c$, by using  a cylindrical geometry,  and using a
mock  slit  geometry,  $N_{cs}$,  that  mimics  typical  spectroscopic
observations and those of our survey (\S~\ref{sec:observations}).  The
mock  spectroscopic   slits  are  rotated   through  several  possible
orientations in the simulation to calculate the possible variations in
observed  pair  fraction  caused   by  the  random  alignment  of  the
spectroscopic slitlets and the orientation  of the galaxy pairs on the
sky.  Our sample of potential LBGs are identified in the simulation by
matching the  two point correlation  function of objects with  a given
minimum  infalling  velocity to  the  observed correlation  functions.
Using lines of sight through  the entire length of the simulation box,
we are able to approximate the projected close pair count of LBGs over
a defined  redshift path.  Finally,  we use multiple  randomly aligned
copies of the  simulation box to explore the full  line of sight depth
of  the observed  sample as  a means  to test  our  simulation results
against the full redshift range of the observations.

%----------------------------------------------------------
\subsection{ Observations}
\label{sec:observations}%Section 2.1
%----------------------------------------------------------

We  design certain  aspects of  the simulation  analysis for  a direct
comparison to  the imaging and  spectroscopic $z\sim3$ LBG  surveys of
C05 and  S03. The survey of  C05 consists of deep  $u'$BVRI imaging of
nine  separate fields  over $\sim$$465$  square arcmin  using  the Low
Resolution Imaging  Spectrometer \cite[LRIS;][]{Oke1995, McCarthy1998}
on  $10$-metre  Keck  I   telescope  and  the  Carnegie  Observatories
Spectroscopic Multislit and Imaging Camera \cite[COSMIC;][]{Kells1998}
on  the   $5$-metre  Hale   telescope  at  the   Palomar  Observatory.
Approximately  800  photometric  LBG  candidates were  selected  in  a
conventional  manner that  uses  their $u'$BVRI  colours.  The  sample
contains  $211$ colour-selected, spectroscopically  confirmed $z\sim3$
LBGs with m$_R \lesssim 25.5$  and a redshift distribution of $\langle
z\rangle=3.02, 1\sigma=0.3$.   The nine fields of  the survey minimize
the effects  of cosmic  variance.  Detailed information  regarding the
colour-selection  technique  and  survey  specifics can  be  found  in
C05. The survey of S03  consists of the publicly available photometric
catalog of  $\sim2500$ $z\sim3$ LBGs  and the the  spatial correlation
results using a spectroscopic subsample of $\sim800$ LBGs.

Although the  sensitivity limits of $8$-metre  class telescopes enable
photometric   detection  of   $z\sim3$  LBGs   to   m$_R  \lesssim27$,
spectroscopic  confirmation is  limited  to those  with m$_R  \lesssim
25.5$ using  reasonable integration times.   The spatial distribution,
or clustering, of the spectroscopic  sample has been used to infer the
average  mass  of  LBGs  in  the  context  of  $\Lambda$CDM  cosmology
\citep[][hereafter C06]{Adelberger05, Cooke06}  and is determined from
the m$_R  \lesssim25.5$ subsample.  For comparison  to our simulation,
we  only consider  LBGs  that have  m$_R  \lesssim 25.5$  in order  to
compile a  sample with ($1$)  accurate photometry ($ <  0.2$ magnitude
uncertainties), ($2$) follow-up  spectroscopic confirmation, and ($3$)
a measured spatial correlation function.

The  C05 survey  is  a conventional  multi-object spectroscopic  (MOS)
survey originally  designed to obtain  a large number of  $z\sim3$ LBG
spectra  to  cross-correlate  with  quasar  absorption  line  systems.
Although it is unclear whether the presence of quasars in these fields
produces a clustering bias for  LBGs near the same redshift range, the
background quasars for  six of the nine fields surveyed  are at a much
higher  redshifts  than the  $\langle  3.0\rangle,  1 sigma=0.3$  LBGs
probed (see C05), thus eliminating any potential clustering bias.  Any
clustering bias for the remaining three fields is likely small because
the LBG  correlation values for the  nine fields in  our survey agree,
within      the     uncertainties,      to     the      results     of
\citet{Adelberger2003,Adelberger05} on  the $17$ field  survey of S03.
Nevertheless, we generate our simulation sample based on the values of
S03 to help  alleviate any potential bias.  Finally,  we note that two
of the serendipitous spectroscopic close pairs in our survey are found
in the  three fields  potentially biased by  the targeted  quasars but
exist  at  much  different   redshifts  as  compared  to  the  quasars
($\delta_z$ corresponding to $>200$  h$^{-1}$ Mpc, comoving) as not to
be biased.

\begin{figure}
\begin{center}
\scalebox{0.25}[0.25]{\rotatebox{0}{\includegraphics{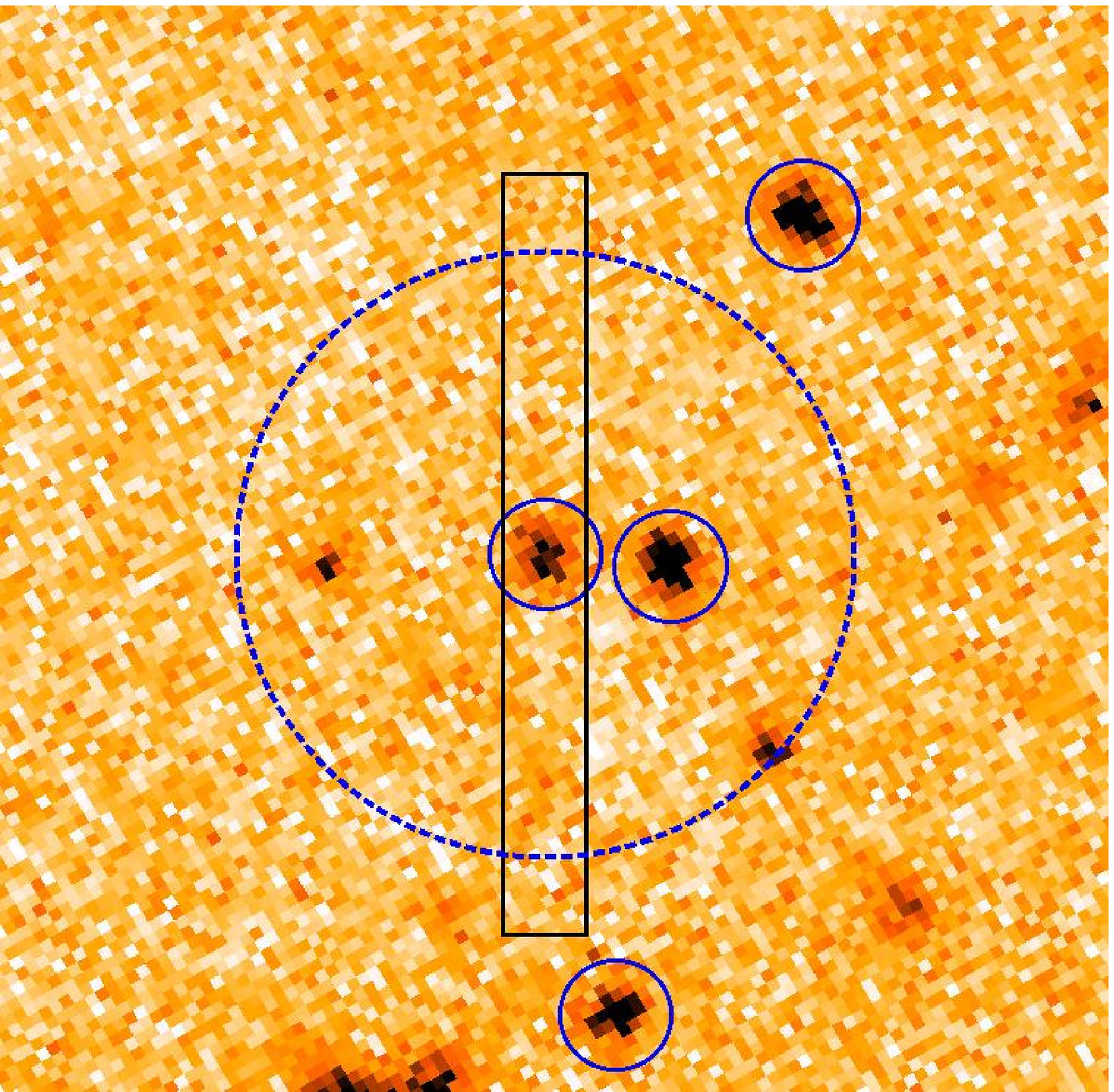}}}
\scalebox{0.25}[0.25]{\rotatebox{0}{\includegraphics{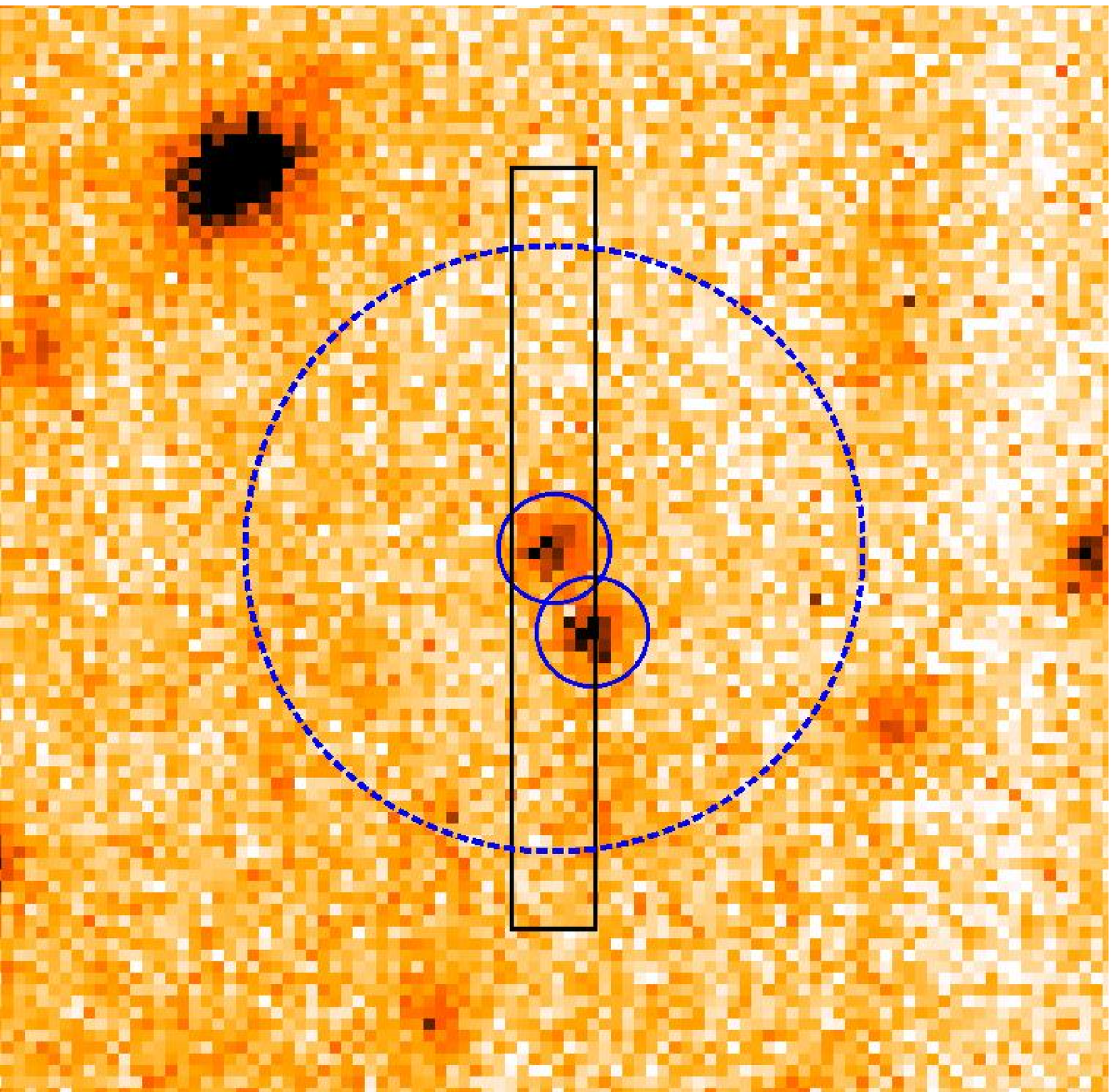}}}
\caption{Illustration  of  conventional  spectroscopic  slit  geometry
  (data  from the  survey  of \citet{Cooke05})  and  mocked-up in  our
  simulation  analysis.    In  both   panels,  the  geometry   of  the
  spectroscopic   slitlets   is   shown   by   the   rectangles,   the
  colour-selected $z\sim3$ LBGs are  marked using small circles, and a
  radius $30 h^{-1}$ kpc at $z=3$ is denoted by a large (dashed) circle
  centred on the targeted LBG.  Although the actual slitlet dimensions
  vary  for  each target,  those  illustrated  here  have the  average
  dimensions of  our survey  {\it (see text)}.   The direction  of the
  spectroscopic  dispersion precludes  acquisition of  objects  to the
  left and right  of the slitlets as depicted in  the two panels.  For
  example  in the  left panel,  the close  companion to  the immediate
  right of the targeted galaxy,  as well as the more distant companion
  to the  upper right  in this highly  clustered case,  cannot receive
  spectroscopy and must await future observations which are not always
  possible.  Occasionally,  the bulk of the  flux of an  LBG pair will
  fall serendipitously  into a  slitlet as shown  in the  right panel.
  Each slitlet  in the survey  is among $\sim30-40$  similarly aligned
  slitlets  acquired  per  telescope  pointing that  are  oriented  to
  minimize atmospheric dispersion at  the time of the observations and
  not designed  to align  with the orientation  of close pairs.   As a
  result, any  LBG pairs that  fall into the slitlets  randomly sample
  the true underlying close pair fraction.}
\label{fig:LBG}
\end{center}
\end{figure}

Conventional MOS  surveys of LBGs  target single LBGs, not  LBG pairs,
and  are designed  to  typically  have the  same  orientation for  the
multiple  slitlets located on  each slitmask.   As such,  the slitlets
have orientations that  are random with respect to  the orientation of
LBG pairs on the sky.  As  a result, the fraction of serendipitous LBG
pairs that fall into the the MOS slitlets enables an accurate sampling
of  the  underlying close  pair  fraction.   An  illustration of  this
concept  for one of  the many  slitlets on  a multiobject  slitmask is
shown  in Figure~\ref{fig:LBG}.  Although  LBGs cluster,  the relative
low surface density of $z\sim3$ LBGs results in very few pairs falling
serendipitously  into the slitlets.   \citet[][hereafter C10]{Cooke09}
identify  $10$ LBGs  in  $5$ serendipitous  spectroscopic close  pairs
($\lesssim20  h^{-1}$ kpc, physical).   The serendipitous  close pairs
provide spectroscopically identified  interacting events to compliment
photometric close  pairs and morphological  classifications which have
previously been the only means to identify high redshift interactions.
Finally, because  the instruments, method, and analysis  of our survey
are virtually identical to most other conventionally acquired surveys,
and specifically  to that of S03,  it is valid to  compare the overall
results from this work.

Typical  $z\sim3$ LBG spectra  have a  signal-to-noise ratio  (S/N) of
only  a few, but  in practice  the strong  UV emission  and absorption
features  and   continuum  profiles  provide  a   means  for  reliable
identification.  Nevertheless,  cautious of  the inherent low  S/N, we
assign  a confidence qualifier  to the  spectroscopic identifications.
For our pair analysis, we  test two samples from the observations: The
full sample of $211$ LBGs and  a sample of $140$ LBGs with the highest
S/N which we term the highest confidence sample.

%----------------------------------------------------------------------
%
%
%
%
%----------------------------------------------------------------------

%----------------------------------------------------------------------
%
%
\begin{figure}
\begin{center}
\scalebox{0.43}[0.42]{\rotatebox{0}{\includegraphics{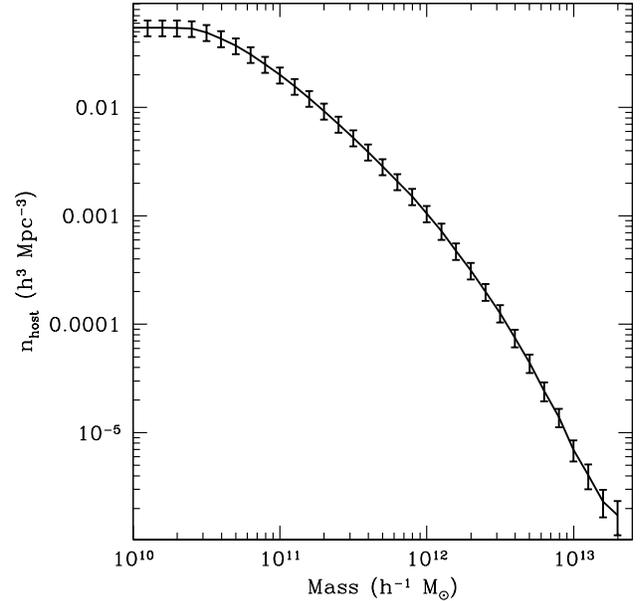}}}
\caption{ The cumulative mass function  of host haloes as derived from
  our $120$~$\hMpc$  simulation box at  $z = 3$.  Error  bars estimate
  cosmic variance using jackknife errors from the eight octants of the
  computational volume.}
\label{fig:massfunc}
\end{center}
\end{figure}
%
%
%----------------------------------------------------------------------

%----------------------------------------------------------------------
%
\begin{figure}
\begin{center}
\scalebox{0.43}[0.42]{\rotatebox{0}{\includegraphics{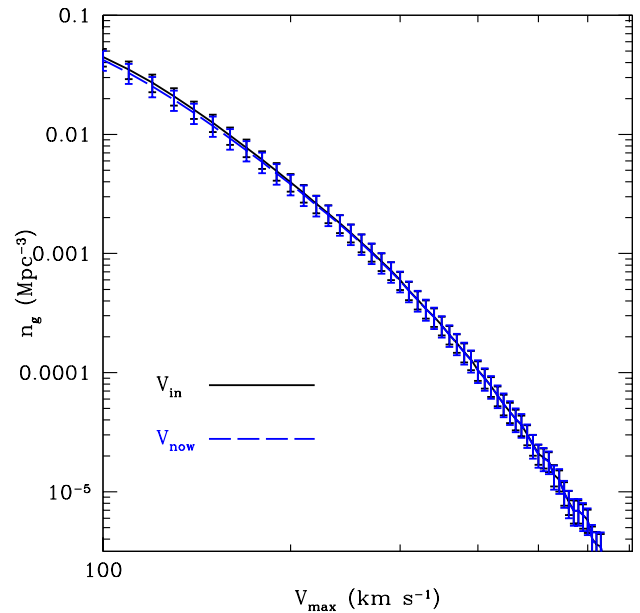}}}
\caption {  The vertical axis  shows the cumulative number  density of
  galaxies in our  simulation catalog as a function  of velocity using
  $V_{in}$  (solid black  line) and  $V_{now}$ (blue  dashed  line) at
  $z=3$ to identify subhaloes as galaxies .  The error bars shown were
  generated  by summing  in  quadrature the  jackknife  error and  the
  realisation-to-realisation  scatter  and  represents errors  due  to
  cosmic variance in the simulation.  }
\label{fig:vmf}
\end{center}
\end{figure}
%
%
%----------------------------------------------------------------------

The color-selection  technique (\citep[e.g.][]{Steidel2003,Cooke05} is
highly efficient  in targeting  $z\sim3$ LBGs and  removing background
and foreground  sources.  The  observed 2-D color-selected  close pair
fractions were  estimated after a correction for  chance alignments by
generating  random catalogs  matched to  the density,  dimensions, and
photometric   selection   functions    specific   to   the   C05   and
\citet{Steidel2003} surveyed fields.

%----------------------------------------------------------
\subsection{ Simulations}
\label{sec:simulations}%Section 2.2
%----------------------------------------------------------

The simulation used for the  large scale structure and host haloes was
performed  using  the Adaptive  Refinement  Tree  (ART) $N$-body  code
\citep{kravtsov_etal:97} for  a universe with a  standard cosmology of
$\Omega_m= 1 - \Omega_{\Lambda} = 0.3$, $h=0.7$, and $\sigma_8 = 0.9$.
The  simulation  followed the  evolution  of  $512^3$  particles in  a
comoving box  of $120 \hmpc$ on a  side, with a particle  mass of $m_p
\simeq  1.07  \times 10^9  \hMsun$.   More  details  can be  found  in
\citet{Allgood05} and \citet{Wechsler05}.  The root computational grid
was comprised of $512^3$ cells and was adaptively refined according to
the evolving local density field to a maximum of $8$ levels.  The peak
spatial  resolution  is   $h_{\mathrm{peak}}  \simeq  1.8  ~\hkpc$  in
comoving units.

In this  simulation the  distinct host haloes  are identified  using a
variation     of     the     Bound    Density     Maxima     algorithm
\citep[BDM,][]{klypin_etal:99}.    In  this   method   each  halo   is
associated with  a density  peak.  This peak  is identified  using the
density field smoothed with a $24$-particle SPH kernel \citep[see][for
  details]{Kravtsov04a}. The halo virial radii and mass are calculated
for the host halo in the simulation box.

The halo virial radius, $\Rvir$, is  defined as the radius of a sphere
whose centre is  the density peak, with mean  density $\Dvir(z)$ times
the mean  density of the universe.  The  virial overdensity $\Dvir(z)$
comes from the spherical  top-hat collapse approximation.  In our case
this     is    computed    using     the    fitting     function    of
\citet{bryan_norman:98}.    The  simulation  assumes   a  conventional
$\Lambda$CDM  cosmology  which  yields  $\Dvir(z=0)  \simeq  337$  and
$\Dvir(z) \rightarrow 178$ at $z \gsim 1$.  The virial mass is used to
characterise  the masses  of distinct  host haloes,  the  haloes whose
centres do not lie within the virial radius of a larger system.

Figure \ref{fig:massfunc} shows  the host halo mass at  $z=3$ from the
procedure  described above.   The  Figure illustrates  host halo  mass
function, complete  to virial masses $M \gsim  10^{11.0} \hMsun$.  The
uncertainties are calculated by  a jackknife error technique. They are
computed by removing one of the eight octants of the simulation volume
and recalculating  the mass function.   These error bars  estimate the
uncertainty in host halo counts from cosmic variance.

The substructures originally located  in these host haloes are removed
and  replaced  by  substructures  generated  using  the  algorithm  of
\citet{Zentner05}.   This analytic  method allows  the  generation and
examination    of     substructure    with    effectively    unlimited
resolution. Each host  halo in this simulation catalog  has a randomly
generated  mass accretion history  using the  extended Press-Schechter
formalism    \citep{Bond91,LC93}    with    the   implementation    of
\citet{Somerville99}.

Once these mass accretion histories and merger trees are generated, we
track  the history  of  the new  subhaloes  as they  evolve.  As  each
subhalo merges into the host  it is assigned an initial orbital energy
and angular  momentum.  Then the  routine calculates the orbit  of the
subhalo  inside a potential  from the  host halo  between the  time of
subhalo accretion  to the epoch  of observation.  Tidal mass  loss and
dynamical  friction are  modeled  to determine  the  effects of  these
interactions on the  mass of the subhalo.  The  halo's density profile
is  modeled  using  the  the \citet[][NFW]{NFW97}  profile  with  halo
concentrations set according  to the algorithm of \citet{wechsler:02}.
Finally,  all  subhaloes  are  tracked until  their  maximum  circular
velocities drop below $\Vmax =  80$~\kms. Haloes which fall below this
threshold  are removed  from the  simulation.  This  step is  to avoid
excess computing  time calculating the small, tightly  bound orbits of
objects that are  not likely to host a luminous  galaxy.  We refer the
interested reader to \S~3 of Z05 for the full details of this model.

This process  is repeated four times  for each host  halo to determine
the effects of variation in  the subhalo populations with a fixed host
halo population.  In addition  to generating substructure catalogs for
each separate ``realisation'' of  the model we perform three rotations
of each simulation volume.   These rotations provide us with different
lines  of sight  through  the substructure  of  the simulation.   This
provides a total of $12$ effective realisations for us to gather close
pair statistics.

Z05 demonstrates that this method is successful at reproducing subhalo
count  statistics,  radial  distributions,  and  two-point  clustering
statistics  measured in  high-resolution $N$-body  simulations  in the
regimes  we  use here.   This  model's  results  agree with  numerical
treatments over $3$ orders of magnitude, or more, in host halo mass as
well as  a function  of redshift.  Moreover,  this technique  has also
proved  useful  in generating  close  galaxy  pair  counts that  match
observations    in    the   local    universe    to    $z   \sim    1$
\citep{Berrieretal06}.

In  addition  to our  primary  simulation  sample  described above,  a
subsample  of  dark matter  haloes  from  the Millennium-2  simulation
\citep[see][for more details]{MBK2009} are used to test our methods in
a pure $N$-body simulation.

%----------------------------------------------------------
\subsection{Assigning Galaxies to Haloes and Subhaloes}
\label{sec:model}%Section 2.3
%----------------------------------------------------------

After  computing the  properties of  haloes and  subhaloes in  a \lcdm
cosmology, the next  step is to map galaxies on  to these objects.  We
use the maximum circular velocity that  the subhalo had at the time it
was accreted into the host halo,  $\Vin$, to define the objects in our
sample.  The choice  of $\Vin$ mimics a case where  a galaxy is highly
resistant  to baryonic  mass loss  when  compared to  its dark  matter
halo. As such,  the luminosity of the galaxy is  unchanged by the loss
of matter  due to  tidal interactions.  This  case assumes a  model in
which the  luminosity of  a galaxy is  set in  the field and  does not
change after merging into the  host system.  Thus we assume that there
is a  monotonic relationship between halo  circular velocity, $\Vmax$,
and galaxy  luminosity.  This model  does not account for  any effects
which  might alter the  galaxies intrinsic  luminosity or  which might
interfere  with  observations,   such  as  galaxy-galaxy  interactions
triggering enhanced  star formation  or dust obscuration.   In effect,
this model assumes a perfectly  observable universe with a strong halo
assumption that  galaxy properties  are set by  the dark  matter halos
they reside  in. The results  of \citet{Conroy2006} suggest  that this
form  of SHAM  may be  used to  examine the  clustering  statistics of
$z=2-4$ LBGs.   Recent works have  suggested that it is  reasonable to
assume  a   halo-UV  luminosity  relation   (essentially  a  halo-star
formation    rate   relation)    at   the    redshifts    we   examine
\citep{Simha2010,Conroy2008,Stewart2009}.   We discuss the  effects of
this    method    of   halo    assignment    on    the   results    in
\S~\ref{sub:Discussion}.

In addition to  the $V_{in}$ model, a second toy  model is tested that
uses the $V_{max}$  of all haloes at the epoch  they are observed.  We
refer to  this model as the  $V_{now}$ model.  This  model describes a
physical  scenario in  which  the dark  matter  and luminous  baryonic
matter are  stripped from subhaloes proportionally.  This  is in stark
contrast  to  the  $V_{in}$  model  where  the  luminous  baryons  are
resistant to mass loss.  This second model has proved to be inadequate
in reproducing  close pairs of  galaxies and features observed  in the
two point correlation function of galaxies locally, but is tested here
for the purposes  of completeness.  Although baryons are  likely to be
stripped from a halo, it is unlikely that they will be stripped at the
same rate as the dark matter.   Again, this model does not account for
the possibility of enhanced star formation due to galaxy interactions.

Figure   \ref{fig:vmf}  shows   the  cumulative   number   density  of
``galaxies'' identified  in our simulations,  $n_g$, as a  function of
their maximum  circular velocities.  The black solid  line shows $z=3$
galaxies using  $\Vin$ as an  identifier, while blue dashed  line uses
$V_{now}$ to  generate the  function. Our catalogs  are complete  to a
$V_{max} = 100 $ km s$^{-1}$.

\begin{figure}
\begin{center}
\scalebox{0.43}[0.43]{\rotatebox{0}{\includegraphics{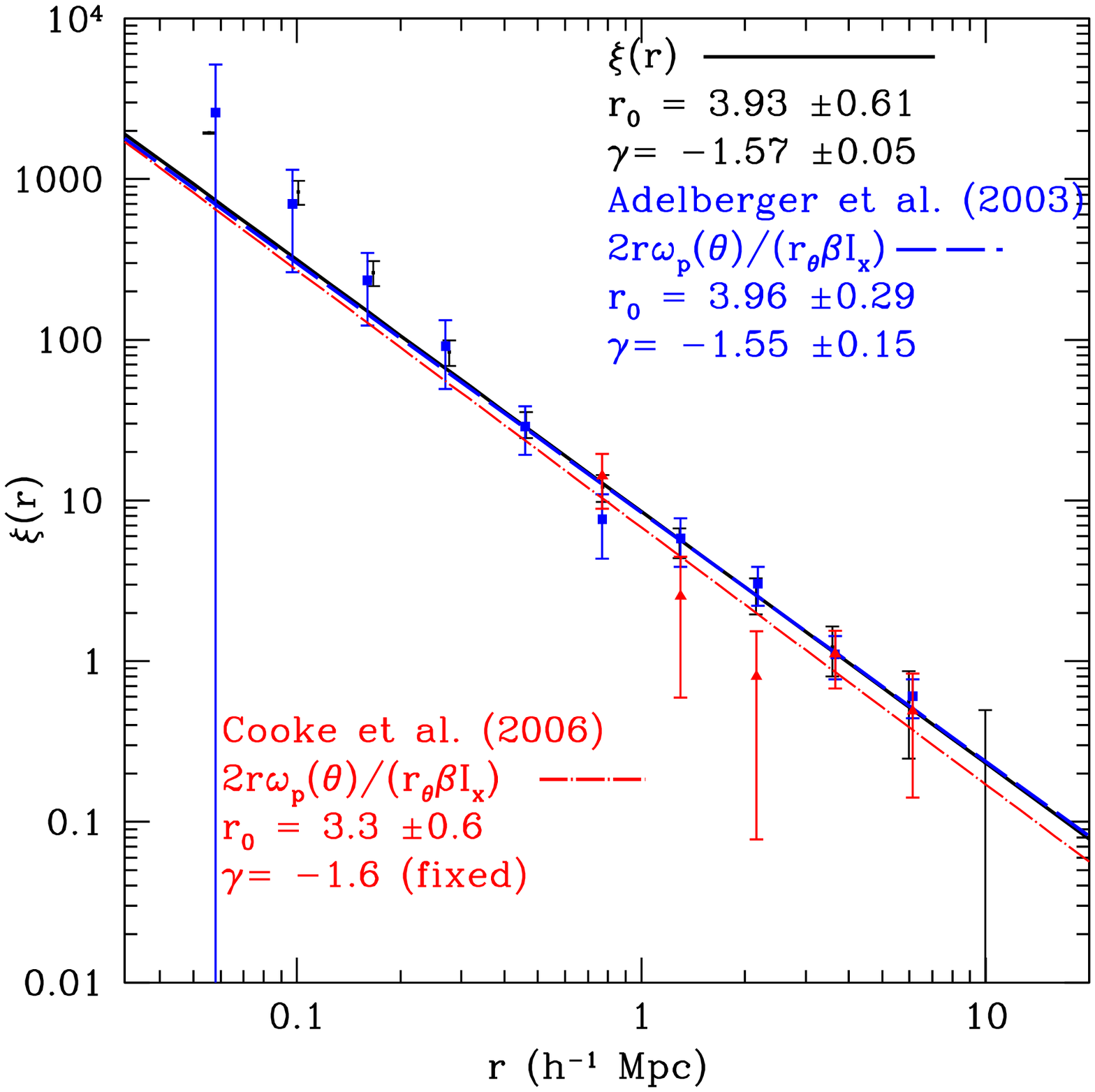}}}
\caption { The three dimensional two-point correlation functions.  The
  blue dashed  line, and square points,  are the real  space two point
  correlation  function  measured from  $z\sim3$  LBG observations  of
  \citet{Adelberger2003}.   The red  dot-dashed  line, and  triangular
  points, are  the real space two point  correlation function measured
  from  $z\sim3$  LBG  observations  \citet{Cooke06}.   Both  sets  of
  observations  are converted  from angular  correlation  functions to
  three      dimensions     using      the      approximation     from
  \citet{Adelberger2003}.   Here the  $\beta$ and  $I_x$ are  the Beta
  function and the  incomplete Beta function. The solid  black line is
  the spatial two point  function generated from the four simulations.
  The  uncertainties  are generated  by  a  combination of  jack-knife
  errors and  realisation to  realisation scatter.  Using  haloes with
  $V_{in} \ge 133$ km s$^{-1}$  in our simulations results in the best
  match to the observations.  }
\label{fig:2pt}
\end{center}
\end{figure}

In addition to testing a standard  SHAM sample and in order to make as
direct a comparison as possible  with observational data, we match the
two-point spatial  correlation functions of our  model galaxy catalogs
to the observed $z \sim  3$ LBG two-point spatial correlation function
of  \citet{Adelberger2003}.  The  correlation functions  are  shown in
Figure  \ref{fig:2pt}.  We  note that  the low-resolution  spectra and
intrinsic star forming processes of $z\sim3$ LBGs make it difficult to
obtain  precise redshifts  from the  emission and  absorption features
\citep{Shapley2003,Adelberger2003}.   We consider  the  effect of  LBG
redshift uncertainties on the pair fractions in the next section.  The
spatial  correlation functions  of  \citet{Adelberger2003} used  here,
incorporate  LBG angular  information  and adopt  a prescription  (see
Appendix C  of that  work) that aims  to minimize redshifts  errors in
order to estimate the true three-dimensional correlation function.  We
fit  the  region  from  approximately  the  innner  separation  radius
computed by  that prescription  out to higher  radii and  thus heavily
dominated by the two halo term region.

The  resulting three dimensional  two-point correlation  function from
the  simulations  is averaged  over  all  four  catalogs, with  errors
including  realisation-to-realisation scatter  and jack  knife errors.
Both the fit  to the observed ``real space''  correlation function and
the simulation follow a power law of the form:
\begin{equation}
\xi(r)=(r/r_0)^{-\gamma},
\label{eqt:2pt}
\end{equation}
where  $r_0$ is  the spatial  correlation length  and $\gamma$  is the
power law slope.  The analysis of C06 places the correlation length at
$r_0 = 3.3 \pm 0.6$ using  a fixed $\gamma = 1.6$, whereas the results
of \citet{Adelberger2003}  find a  value of $r_0  = 4.0 \pm  0.6$ with
$\gamma  = 1.57 \pm  0.14$ for  the larger  S03 survey  dataset.  When
matching the  two-point correlation function of the  simulation to the
data we  find that haloes with  $V_{in} \ge 133$  km s$^{-1}$ best-fit
the parameters  of the observations.   These haloes produce  values of
$r_0 =  3.93 \pm 0.61$  and $\gamma =  1.57 \pm 0.05$.   The $V_{now}$
model that best matches the  observed correlation function is found to
have $V_{now} \ge  142$ \kms with $r_0 = 3.99 \pm  0.64$ and $\gamma =
1.54 \pm 0.05$.

As discussed above, the  spatial distribution, or clustering, has been
used to infer the average mass  of LBGs in the context of $\Lambda$CDM
cosmology.   \citet{Adelberger05} and C06  find the  mass of  the m$_R
\lesssim  25.5$  LBG  spectroscopic  sample to  be  $\langle  M\rangle
\sim10^{11.6 \pm 0.3} M_\odot$.  The  mean mass of our sample compares
well at $\langle M\rangle = 10^{11.54} \hMsun$.

%----------------------------------------------------------------------
%
%
%
%
%---------------------------------------------------------------------

%----------------------------------------------------------
\subsection{Defining Close Galaxy Pairs} 
\label{sec:nc}%Section 2.4
%----------------------------------------------------------

Now that we  have selected our candidate LBG  haloes we must determine
the  best  way to  ``observe''  our sample  in  order  to make  direct
comparisons with observational  results.  We examine spectroscopically
discernible LBG pairs first.

We define a  spectroscopic close pair three ways.   Our first criteria
include galaxies  with a separation of $10  - 30 ~\hkpc$ on  the sky a
relative velocity  difference in the  range of $-500 \le  V_{diff} \le
500$  km  s$^{-1}$.   The  separation  range on  the  sky  reflects  a
conventionally  determined distance  in which  close galaxy  pairs are
considered merger  candidates and is  designed to exclude  close pairs
that  would likely  appear  as a  single  galaxy in  the images.   The
velocity  difference,  if assumed  to  be  dominated  by the  peculiar
velocities of the galaxies, corresponds to the haloes that have a high
probability  of merging.   The $10  -  30 ~\hkpc$  separation is  well
measured  by our  data.   The  compact, near  point  source nature  of
$z\sim3$  LBGs  and the  typical  FWHM  seeing  of the  images  allows
separation of individual  galaxies down to $\sim6 h^{-1}$  kpc and the
average length of the  slitlets used in the spectroscopic observations
is  $\sim 37.7  ~\hkpc$ at  $z =  3$, with  the bulk  of  the slitlets
probing beyond the conventional maximum separation.

The spectroscopic FWHM resolution  of our observations is $\sim400$ km
s$^{-1}$ but velocity  offsets as small as $\sim  200$ km s$^{-1}$ can
be  measured  using   multiple  cross-correlated  spectral  lines  and
high-significance Ly$\alpha$ features.  As  a result, we test a second
criteria  that  includes  pairs  with  $0-10  ~\hkpc$  separations  in
projection on the sky for objects that may appear indistinguishable in
the images but  show clear indications of two  separate spectra with a
sufficiently  large  velocity   difference  ($V_{diff}=  \pm  200$  to
$\pm500$ km s$^{-1}$).

Finally, we exploit the  behaviour of the prominent Ly$\alpha$ feature
in $z \sim 3$ LBGs which can be observed in absorption, emission, or a
combination of both.  The peak of this feature in emission is observed
to  be  redshifted from  the  systemic  redshift  by $450\pm300$  \kms
\citep{Adelberger2003},  with the tail  of the  distribution extending
beyond $1000$  \kms.  The observed redshifted peaks  are attributed to
galactic-scale outflows  driven by  stellar and supernovae  winds.  In
this  picture, the  blue wing  of the  Ly$\alpha$ emission  feature is
absorbed  by neutral  gas moving  toward the  observer  and Ly$\alpha$
photons traveling away from  the observer are shifted off-resonance as
they  scatter off  receding portions  of the  outflow,  enabling their
escape back toward the observer.

C10 find that every LBG in the serendipitous spectroscopic close pairs
in  their sample and  every spectroscopic  LBG with  a colour-selected
close  ($\lesssim 20  ~\hkpc$)  LBG exhibits  Ly$\alpha$ in  emission.
Because  the   Ly$\alpha$  feature  is  typically   detected  at  high
significance, we test a third  criteria that includes close pairs with
no minimum  separation on the sky  and with no  $V_{diff}$.  The large
velocity  dispersion   Ly$\alpha$  peaks   can  help  to   enable  the
identification  of  LBG  pairs  with  little  or  no  actual  velocity
difference.   Random  samplings of  the  Ly$\alpha$ emission  velocity
offsets   show  that   nearly  all   such  galaxy   pairs   should  be
spectroscopically discernible.

To summarise we  test three different criteria to  select close pairs.
Each of these criteria use a maximum outer radius of $30 ~\hkpc$ and a
maximum velocity  difference of  $\pm500$ km s$^{-1}$.   The remaining
parameters for the different criteria are:
\begin{itemize}
\item (A)  Pairs with  minimum separations less  than $10  ~\hkpc$ are
  always excluded  (our fiducial sample  for the cylinders).   This is
  designated as $N_c$ for  cylinders and $N_{cs}(A)$ for spectroscopic
  slits.
\item  (B) Pairs with  minimum separations  between $0-10  ~\hkpc$ are
  included  (thus  pairs  with   separations  $0-30  h^{-1}$  kpc  are
  considered)  if there  velocity  difference is  $V_{diff} \ge  200$.
  These results are labeled $N_c(B)$ for cylinders and $N_{cs}(B)$ for
  spectroscopic  slits. This case  allows us  to test  an intermediate
  case between criteria (A) and (C).
\item (C) No minimum separation or velocity difference. All pairs with
  separations $<30 h^{-1}$ kpc and $-500 < V_{diff} < 500$ km s$^{-1}$
  are  identified.    This  will  be  our  fiducial   sample  for  the
  spectroscopic  slit measurements, $N_{cs}$.   These are  reported as
  $N_c(C)$ for cylinders and $N_{c}$ for spectroscopic slits.
\end{itemize}
The differences in the results  of these three criteria are small (see
\S~\ref{sub:evolution}).

With  these  parameters  we  calculate  the  close  pair  fraction  of
galaxies. This quantity is defined as:
\begin{equation}
N_c \equiv \frac{2 n_{\rm p}}{n_g}.
\label{eqt:nc}
\end{equation}
Here  $n_{\rm p}$  is the  number density  of pairs  and $n_g$  is the
number density of galaxies in  the sample volume.  Thus $N_c$ reflects
the fraction of galaxies that have close companions.

%----------------------------------------------------------
\subsection{Spectroscopic Companions}
\label{sec:speccomp}%Section 2.5
%----------------------------------------------------------

We can perform an analysis of the underlying close pair fraction using
the serendipitous spectroscopic LBG pairs  of C10 and by mimicking the
observation approach  of these  data in the  simulation.  In  order to
best   determine  the   probability  of   observing   a  serendipitous
spectroscopic  pair  in  the  simulation,  we generate  slits  in  two
specific  ways.  Firstly, we  construct mock  slits having  the actual
lengths and widths used in the observations with the objects placed at
the observed locations in the  slitlets.  We place a randomly assigned
slit on the candidate LBG haloes in the simulation and then rotate the
slit through $360$  degrees, in steps of $20$,  to compute the average
$N_{cs}$  for all  pairs ``observed''.   Within the  slit  geometry we
utilise criteria (C) above  as our fiducial sample.  Measurements made
in these randomly assigned slits are designated $N_{cs}(R)$.

Secondly,  we   recompute  $N_{cs}$  from  our   simulations  using  a
prototypical   slit   length  and   width   to   make  a   generalized
``observation''.  The prototypical slit  has a width of $1.37$ arcsec,
the mean width of the slits used in the observations, for a half width
of $3.7 ~\hkpc$.  The prototypical slit length is $60.0 ~\hkpc$, which
is $11.13$ arcsec at $z=3$.  All lengths are given in proper, physical
units.   The   $N_{cs}$  measured  in  the   spectroscopic  slit  only
incorporates pairs observed in one  of the rotations of the slit. Thus
the $N_{cs}$ is  smaller then the total $N_c$ for  any of the criteria
discussed above. Because  the candidate LBG haloes are  centred in the
prototypical slits we  only need to perform $10$  rotations over $180$
degrees to  determine the random  slit count.  Again, we  use criteria
(C) above as our fiducial  sample.  Modeling the observations in these
two ways enables a direct  comparison to the C10 analysis and provides
results  that   can  be  applied  in   a  more  general   way  to  any
conventionally acquired survey.

%---------------------------------------------------------------------
%
%
\begin{figure}
\begin{center}
\scalebox{0.43}[0.43]{\rotatebox{0}{\includegraphics{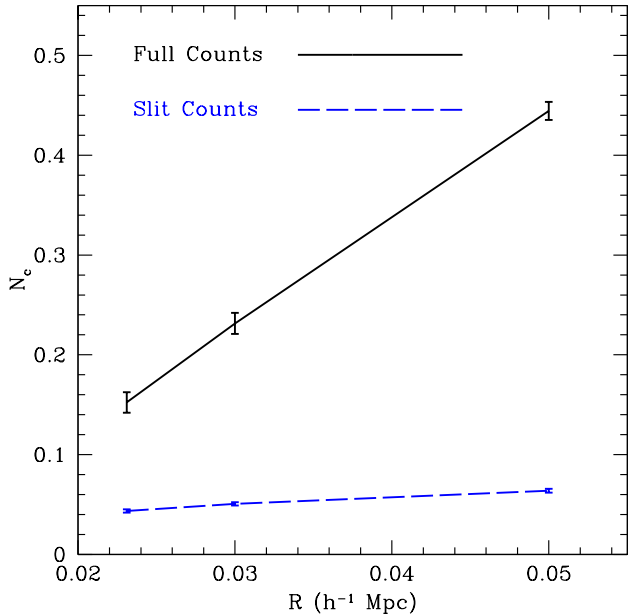}}}
\caption{ The  dependence of  the close pair  fraction on  the maximum
  separation between  the central galaxy and its  companion.  All data
  is calculated  at $z=3$. The solid  black line is  the full cylinder
  sample.  The  blue dashed  line is the  close pair fraction  for the
  spectroscopic  slit  sample.  Uncertainties  are  calculated from  a
  combination of the standard  deviations of the four realisations and
  jack knife errors in the individual simulation boxes.}
\label{fig:RadNc}
\end{center}
\end{figure}
%
%
%---------------------------------------------------------------------

We may also  characterise the dependence of the  sample on the maximum
pair   separation  used.    As  observed   in  Figure~\ref{fig:RadNc},
increasing the maximum radius used for our close pair counts increases
our resulting  $N_c$.  Though the  increase is comparatively  small in
the spectroscopic  slit, we can see in  Figure~\ref{fig:RadNc} that it
is significant for the full cylinder sample.

%----------------------------------------------------------
\subsection{Photometric Companions}
\label{sec:photcomp}%Section 2.6
%----------------------------------------------------------

We make one  other measurement in this work.   We examine the apparent
angular,  or ``photometric'',  close  pair fraction,  $N_p$, and  it's
impurity.  A photometric close pair is defined as one in which we have
no cut on  velocity, and also no pairs are  allowed within the minimum
radius.  This  is essentially a single  line of sight  through the box
designed to  mimic the companion counts  observed in the  plane of the
sky in  photometric surveys that  utilise simple LBG  colour selection
criteria at $z \sim 3$.

Our  simulation identifies  close  pairs within  a  redshift range  of
$\delta z \sim 0.18$ at $z=3$.  The criteria of S03 and C05 select LBG
populations  with $\langle  z\rangle =  3.0, 1\sigma=0.3$.   While our
simulation samples  $\sim 25 $ per  cent of the LBGs  detectable in $z
\sim 3$ surveys, it probes a large enough redshift path to distinguish
objects that are, and are  not, physically clustered.  This is true in
part  because clustering  effects are  negligible beyond  a  radius of
$\sim 10 \hMpc$.  Thus we can  use our analysis to provide an estimate
of $N_c$ and the sample impurity.  Our results for this work are found
in section~\ref{sub:pred}.

%----------------------------------------------------------
\subsection{Diagnosing the effects of Interlopers} 
\label{sec:I}%Section 2.7
%----------------------------------------------------------

Our method  characterises the probability of  chance projections being
identified as  a companion galaxy.   We define the sample  impurity as
the fraction  of "observed" close  pairs in the simulation,  using the
criteria  described above,  that do  not reside  inside a  mutual dark
matter  halo and  are not  a physically  interacting pair.   The total
sample impurity is given by
\begin{equation}
I \equiv \frac{n_{\rm f}}{n_c}.
\label{eqt:impurity}
\end{equation}
Here $n_{\rm  f}$ is the number  density of false galaxy  pairs in the
sample volume  and $n{\rm_c}$  is the number  density of  galaxy pairs
observed in the sample.

\begin{figure}
\begin{center}
\scalebox{0.43}[0.43]{\rotatebox{0}{\includegraphics{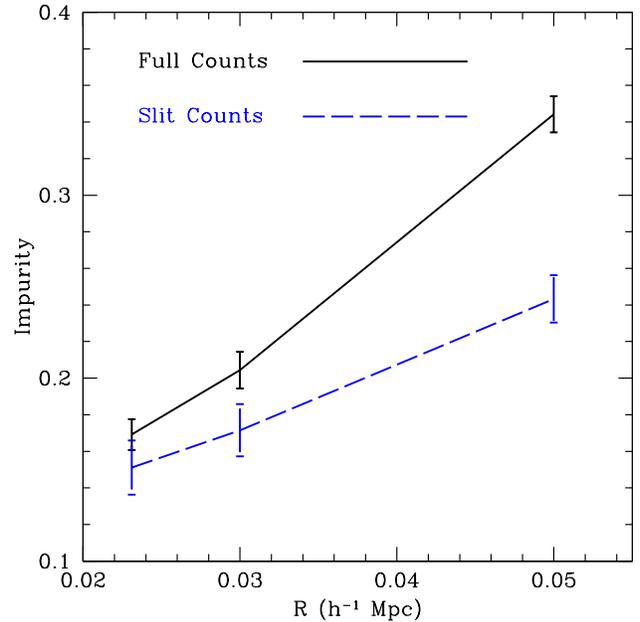}}}
\caption{ The  measured sample impurity using a  varying outer radius.
  The  solid black  line  shows  the impurity  fraction  for the  full
  cylinder sample, while  the blue dashed line shows  the impurity for
  the randomly oriented  spectroscopic slit sample.  Uncertainties are
  calculated  from a  combination of  the standard  deviations  of the
  realisations  and jack  knife  errors in  the individual  simulation
  boxes.}
\label{fig:RadI}
\end{center}
\end{figure}

Figure~\ref{fig:RadI}  identifies the  sample impurity  as  it evolves
with the radii of the cylinder used.  In this case we see how impurity
is effected by  the maximum size of the cylinder. This  is of course a
simple  relationship.  As  we increase  our maximum  radius we  have a
greater chance of identifying a pair of companion galaxies, but also a
greater  risk of  picking up  a  chance projection  of two  physically
unassociated galaxies. In this figure the uncertainties are calculated
by  summing in quadrature  an error  associated with  cosmic variance,
calculated    by    a    jack    knife    error    method,    and    a
realisation-to-realisation scatter.

%----------------------------------------------------------------------
%
%
%
%
%----------------------------------------------------------------------

\begin{figure}
\begin{center}
\scalebox{0.43}[0.43]{\rotatebox{0}{\includegraphics{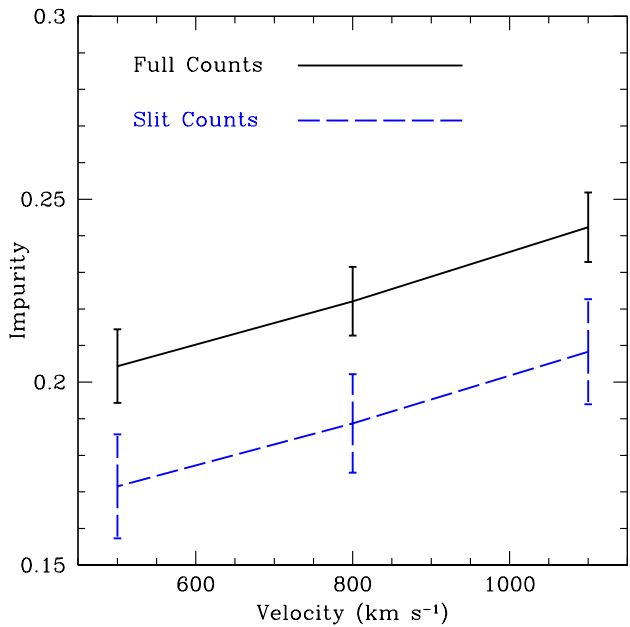}}}
\caption{ The  measured sample  impurity using various  velocity cuts.
  The solid  black line indicates the  impurity of the  sample for the
  full cylinder counts.  The dashed blue line is the impurity recorded
  for the randomly  oriented spectroscopic slit sample.  Uncertainties
  are calculated from a combination  of the standard deviations of the
  four realisations and jack knife errors in the individual simulation
  boxes.    Interestingly  this   demonstrates   that  large   maximum
  velocities do  not significantly  increase sample impurity  over the
  range shown here.}
\label{fig:VelI}
\end{center}
\end{figure}

We also examine the effect that extending the maximum velocity between
galaxies  has  on  both  the  pair fraction  and  the  sample  purity.
Figure~\ref{fig:VelI}  shows this  relationship.  As  before,  using a
larger  velocity  cut increases  both  our  pair  fraction and  sample
impurity.   The  error bars  are  calculated  in  the same  manner  as
Figure~\ref{fig:RadI}.

In  the observational  samples  of S03  and  C05, the  color-selection
techniques are highly efficient ($\gtrsim$90\% effective) in targeting
$z\sim3$  LBGs  and  removing  background and  foreground  sources  as
determined by the spectra.  The observed 2-D color-selected close pair
fractions  reported below  were corrected  for impurity  by generating
random catalogs  matched to  the density, dimensions,  and photometric
selection functions  specific to the  C05 and S03 surveyed  fields and
computing the  fraction of  random unassociated pairs  occurring within
the projected separations for the criteria described above.

%---------------------------------------------------------------------
%
%
%
%
%---------------------------------------------------------------------

%----------------------------------------------------------
\subsection{Galaxy Number Density Problem}
\label{sub:NG}%Section 2.8
%----------------------------------------------------------

For our $z=3$  comparisons we adopted a sample  of potential LBGs that
best describes the data and has $V_{in} \ge 133$ km s$^{-1}$, that is,
the $V_{max}$ of a subhalo when  it is initially accreted into a host,
or simply  the $V_{max}$ of the  object if the halo  is an independent
host  halo.   In  addition  to  matching  the  two  point  correlation
function, these models produce  a comoving number density of galaxies,
$n_g$, that  may also  be compared to  the observational  sample.  The
volumetric  comoving   number  densities   for  the  two   models  are
$n_g(V_{in} \ge 133$ km s$^{-1}) = 0.019 \pm 0.003$ h$^{3}$ Mpc$^{-3}$
and $n_g(V_{now} \ge  142$ km s$^{-1}) = 0.01442  \pm 0.00003$ h$^{3}$
Mpc$^{-3}$.

The comoving number densities for more massive haloes are: $n_g(V_{in}
\ge  200$ km  s$^{-1}) =  0.0039 \pm  0.0006$, h$^{3}$  Mpc$^{-3}$ and
$n_g(V_{in}  \ge  300$ km  s$^{-1})  =  0.00059  \pm 0.00010$  h$^{3}$
Mpc$^{-3}$ .   While we  see that  the $V_{in} \ge  200$ sample  has a
better match to  the number density of $\sim  0.004 \pm 0.002$ h$^{3}$
Mpc$^{-3}$ from LBG observations (see below for further details), this
sample does not match the observed correlation function and produces a
significantly lower value for $N_c$.

The  matched  $V_{now}$  sample  has  a surface  density  in  comoving
coordinates  of $6.14$ h$^{2}$  Mpc$^{-2}$, $10.2$  galaxies arcminute
$^{-2}$,  compared  to   $8.2$  h$^{2}$  Mpc$^{-2}$,  $13.6$  galaxies
arcminute  $^{-2}$ for  the  $V_{in}$ sample.   Both  of these  values
exceed  the $\sim  1.7$ galaxies  arcminute $^{-2}$  from observation.
These  surface densities  are measured  from  the length  of the  $120
\hMpc$ box at $z=3$, a redshift  range of $\delta z \sim 0.2$ and then
corrected for  the total redshift  pathlength observed in  the surveys
and the efficiency of LBG detection as a function of redshift from the
colour selection (assumed  to be $100$ per cent  efficient near $z=3$,
e.g., no lost galaxies as a result of bright stars, low S/N regions on
the chip,  colour detection efficiency, etc.,  as is the  case for the
observations).

Previous  research  has uncovered  similar  discrepancies between  the
number  density  of  matched  massive  haloes and  that  observed  for
$z\sim3$    LBGs     using    different    types     of    simulations
\citep[e.g.,][]{Dave2000,Ouchi04,Nagamine05,Lacey2011}.  Nevertheless,
each propose that the excess  $n_g$ can be resolved by including other
types of high redshift objects,  such as dust obscured and/or low star
formation  rate  galaxies  and  damped Ly$\alpha$  absorption  systems
(DLAs).   The model  of  \citet{Lacey2011} suggest  that without  dust
extinction  the number  of LBGs  would be  $\sim 5$  times  the number
observed.  This  model also recreates  the properties of  the observed
sample once dust  extinction is included. Thus, we  report our results
using the full (high-density) sample  below and explore the effects of
the density mismatch on the results in section~\ref{sub:Density}.

%*********************************************************************

%----------------------------------------------------------
\section{Results}
\label{sec:results}%Section 3
%----------------------------------------------------------

Using the  methods described in  Section~\ref{sec:model} we calculate:
($1$)  the  close  pair   fraction  observed  serendipitously  in  the
spectroscopic  slits,   $N_{cs}$,  ($2$)  the   $10-30  ~\hkpc$  total
spectroscopic close  pair fraction, $N_c$,  ($3$) the $  10-30 ~\hkpc$
photometric close pair fraction, $N_p$.

As  a reminder,  all    three sample  criteria  include  pairs    with
separations $<30$  \kpc and  a velocity  difference  of $\pm500$ \kms.
However, the differences are that criteria (A) excludes all pairs with
separations  $<10$ \kpc, criteria  (B) includes pairs with separations 
$<10$ \kpc if their velocity  difference is $>200$ \kms, and  criteria 
(C) includes pairs with separations  of $<10$ \kpc with no restriction
on the velocity difference.

%----------------------------------------------------------
\subsection{Spectroscopic Close Companion Counts at z=3}
\label{sub:evolution}%Section 3.1
%----------------------------------------------------------

We first note that the observations  of C10, to which we are comparing
these   results,  find   a  serendipitous   close  pair   fraction  of
$N_{cs}=0.071 \pm  0.023$ in the  spectroscopic slits for  the highest
confidence subsample ($140$ LBGs) and $N_{cs}=0.047 \pm 0.015$ for the 
full  sample  of  $211$  LBGs.   In  our  simulations  we  ``observe''
serendipitous close  pairs using the three different  sets of criteria
discussed in \S~\ref{sec:nc}.

In  our  randomly selected  slit  length  sample  (lengths and  object
positions matched to the survey of C05) we find $N_{cs}(R)(A) = 0.0220
\pm 0.0234$,  $N_{cs}(R)(B) = 0.0300 \pm 0.0178$,  and $N_{cs}(R)(C) =
0.0430  \pm 0.0145$.  All  three results  are consistent  within their  
uncertainties.

For  the  prototypical  slit  length  of $60.0  ~\hkpc$,  we  find  an
``observed'' $N_{cs}(A) = 0.0293  \pm 0.0013$, $N_{cs}(B) = 0.0375 \pm
0.0016$, and $N_{cs}(C) = 0.0506 \pm 0.0017$.

As before,  the uncertainties are  a combination of jack  knife errors
from  cosmic variance  and realisation-to-realisation  scatter.  These
measurements reflect our uncorrected  sample and include galaxies that
meet  our velocity criteria  and are  observed as  close pairs  due to
projection effects even  though they do not reside  in the same parent
halo.

Finally, the full observable pair  fraction in the cylinder is $N_c(A)
= 0.228  \pm 0.006$, $N_c(B) =  0.246 \pm 0.006$, and  $N_C(C) = 0.275
\pm 0.007$. As expected, criteria  (B) and (C) produce slightly higher
values,  but they are  not significantly  different from  our fiducial
value (A).

We report  the results of criteria  (A) for the  cylinder and criteria
(C) for  the spectroscopic slits  because criteria (A) is  designed to
match  the  morphological and  close  pair  fraction  analyses in  the
literature and criteria (C) is  best matched to the methodology of the
spectroscopic slit analysis of C10.

\begin{figure}
\begin{center}
\scalebox{0.43}[0.43]{\rotatebox{0}{\includegraphics{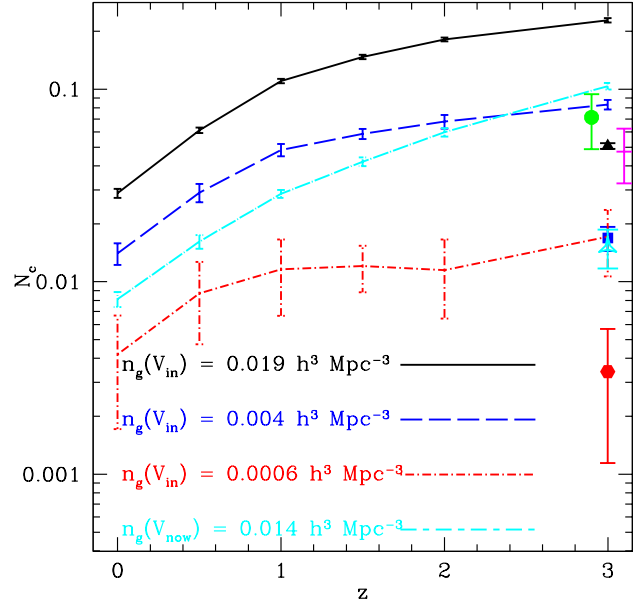}}}
\caption{  Expected spectroscopic close  pair fractions.   Plotted are
  the  expectations from  $z=0-3$  for different  number densities  of
  galaxies corresponding to the specified velocity cuts at $z=3$.  The
  results report  the pair  fraction in cylinders  with radii $r  < 30
  ~\hkpc$  (curves).    The  expected   $N_{cs}$  at  $z=3$   for  the
  serendipitous  pairs  in  "observed"  spectroscopic slits  are  also
  plotted  (points).   The  solid   black  curve  and  triangle  point
  represent  the  close  pair  fraction  of galaxies,  $N_c$,  in  our
  fiducial sample of haloes using the $V_{in} \ge 133$ km s$^{-1}$ cut
  at $z=3$,  and the  same number density  at all  previous redshifts,
  matched  to the LBG  correlation function  observations.  Similarly,
  the  light blue  short-dash long-dashed  curve and  empty triangular
  point represent our fiducial $V_{now} \ge 142$ km s$^{-1}$ sample at
  $z=3$.   More massive  and brighter  samples are  shown  with number
  densities matching the values of  $V_{in} \ge 200$ km s$^{-1}$ (blue
  dashed curve,  square point) and  $V_{in} \ge 300$ km  s$^{-1}$ (red
  dot-dashed  curve, hexagonal point)  at $z=3$.   Uncertainties shown
  for the randomly oriented slits are calculated from a combination of
  jack knife  errors and variance between the  four realisations used.
  Please note that these points  are not corrected for sample impurity
  due to projection effects.  The green dodecagonal point at $z = 2.9$
  and the magenta  dash point at $z=3.1$ represent  the observed $N_c$
  from the high signal to noise sample and the full observed sample of
  $z\sim3$  LBGs \citep{Cooke09},  respectively, and  are  offset from
  $z=3$ for clarity.  }
\label{fig:NC}
\end{center}
\end{figure}

Close pair  fraction predictions  for the full  uncorrected observable
$N_c$    and    the   spectroscopic    slit    may    be   found    in
Figure~\ref{fig:NC}. Note that these values are not corrected for line
of sight projection effects. Here we track objects in the catalog from
$0 \le z \le  3$ that have a number density matched  to a $z=3$ sample
with  $V_{max}$  greater  than  four different  critical  values.   By
holding a  constant $n_g$  cut we  examine the changes  in $N_c$  at a
fixed population size with redshift.  The solid black curve represents
the close pair  fraction for a galaxy sample with  $V_{in} \ge 133$ km
s$^{-1}$ utilising  our standard criteria (A).   The green dodecagonal
point and  the magenta dash point  (offset to $z=2.9$  and $z=3.1$ for
clarity)  are the  observed  serendipitous close  pair  values of  the
highest confidence  and total  $z=3$ sample, respectively.   The solid
black triangle denotes the results from our fiducial slit criteria (C)
and is consistent with both observational samples.

The blue dashed   line and  the red   dot-dashed  line represent  more
stringent  mass cuts with values  of $V_{in} \ge  200$ km s$^{-1}$ and
$V_{in} \ge  300$   km  s$^{-1}$, respectively,   using the   standard
cylinder.   These values are  included to  make  a comparison of $N_c$
with $n_g$,  the (comoving)  number  density of galaxies,  as has been
done in work by other authors, even though their correlation functions
do not closely match the observations.  In  these cases we may examine
the evolution of higher  mass subsamples with  a fixed $V_{in}$,  with
redshift.  The solid  blue square and  the solid red hexagon represent
the expected slit  model $N_{cs}$ for  these higher mass cuts.   As we
can see, they  are both well below the  expected $N_{cs}$ from the LBG
observations.

The  light blue short-long dashed  line  and the  light blue  triangle
represent the value for our $V_{now} \ge 142$  sample. This model also
underpredicts   $N_c$, and does  so even  when  matched  to the number
density    of    galaxies,    $n_g$,    at    all    redshifts    (see
\citet{Berrieretal06} for  more detail).  As  this model  demonstrates
the  same flaws as  our  $V_{in}$  model  and  does not reproduce  the
observed $N_c$  down  to low-$z$,  it  will be  excluded  from further
discussion.

%---------------------------------------------------------------------
%
%
%
%
%--------------------------------------------------------------------- 

At our fiducial slit radius of $30.0 ~\hkpc$ the sample impurity is $I
= 0.1716 \pm 0.0111$ for the slit and $I  = 0.2225 \pm 0.0099$ for the
full cylinder  count.  These values imply  that we have $(1-I)$ $N_c$,
or a real  $N_{cs} \sim 0.042  \pm 0.0015$ (cf.  $N_{cs}  = 0.0506 \pm  
0.0017$, uncorrected)  in the  slits and  $N_c \sim  0.177 \pm 0.005$
(cf. $N_c = 0.228 \pm 0.006$, uncorrected) for the full cylinder.

%---------------------------------------------------------------------
%
%
\begin{figure}
\begin{center}
\scalebox{0.43}[0.43]{\rotatebox{0}{\includegraphics{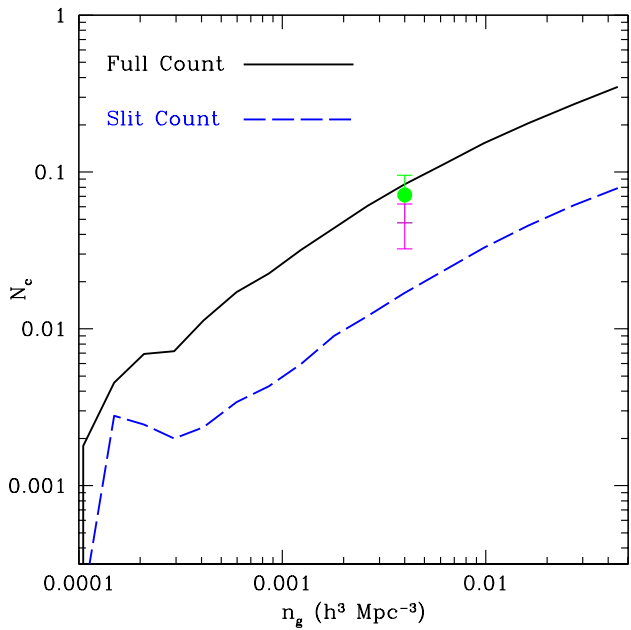}}}
\caption{ Variation  of the $z=3$  close companion count  ($N_c$) with
  galaxy number density ($n_g$).  The black solid line gives the $N_c$
  for the full cylinder while  the blue dashed line gives the $N_{cs}$
  for the randomly oriented  spectroscopic slits.  The green octagonal
  point represents the observed  serendipitous close pair fraction for
  the  high  confidence sample  in  the  spectroscopic  slits, at  the
  observed  number density,  in  the sample  of \citet{Cooke09}.   The
  magenta hash is the pair fraction for the total observed sample.}
\label{fig:z3n}
\end{center}
\end{figure}
%
%
%
%---------------------------------------------------------------------

We   can   generalize    the   results   of   Figure~\ref{fig:NC}   to
Figure~\ref{fig:z3n}.   We examine  the values  of $N_c$  in  both the
spectroscopic  slit (blue  dashed curve)  and the  full  cylinder with
radius  $10 \le  r  \le 30$  h$^{-1}$  kpc (black  solid  curve) as  a
function of the  number density of ``galaxies'' in  the sample, $n_g$.
Here, the observational spectroscopic  slit results seems too high for
the observed  LBG number density  ($n_g \sim 0.004 \pm  0.002$ h$^{3}$
Mpc$^{-3}$) and  appears to be  more consistent with pair  fraction at
the density of  our full $V_{in} \ge 133$ km  s$^{-1}$ sample which is
$\sim5\times$  higher.  What  causes this  inconsistency?   We discuss
this issue in $\S$~\ref{sub:Density}

%----------------------------------------------------------
\subsection{Angular Close Pair Fraction}
\label{sub:pred}%Section 3.2
%----------------------------------------------------------

We  also use  our  simulation  to estimate  the  observed angular,  or
``photometric'',  close  pair  fraction,  $N_p$.   In  this  case,  we
calculate a pair fraction without using a velocity cut and exclude all
objects inside  the minimum projected  radius of $10 h^{-1}$  kpc.  As
discussed above, it  is possible to approximate the  $N_p$ at $z=3$ in
the  simulation  box.  The  simulation  is  $120  \hmpc$ long  in  the
comoving coordinates,  and thus at  $z=3$ has an  approximate redshift
length of  $\delta z =  0.18$ from front  to back.  Due  to clustering
effects being  less pronounced beyond $10  \hmpc$ we may  use this box
for  this purpose,  despite the  redshift range  for the  survey being
$\delta z  \sim 0.6$.  Thus, while  we can easily  determine $N_p$, we
must be cautious that this is only an approximate value.

Figure~\ref{fig:VelI} provides us with some estimate of the effects of
a larger velocity cylinder  on the purity of the  sample.  In the case
of  our  photometric   sample we are  using   an   equivalent velocity
difference based on  the total length of  the box in local Hubble flow
as  well as the  peculiar velocities along the  line  of sight for the
galaxies.  Similarly  we can see   from Figure~\ref{fig:RadI} that the
sample's impurity increases with larger sample radius.  In the case of
$30 ~\hkpc$ this is $\sim 0.20 \pm 0.01$ from the on the sky dimension
of the cylinder alone.

\begin{figure}
\begin{center}
\scalebox{0.43}[0.43]{\rotatebox{0}{\includegraphics{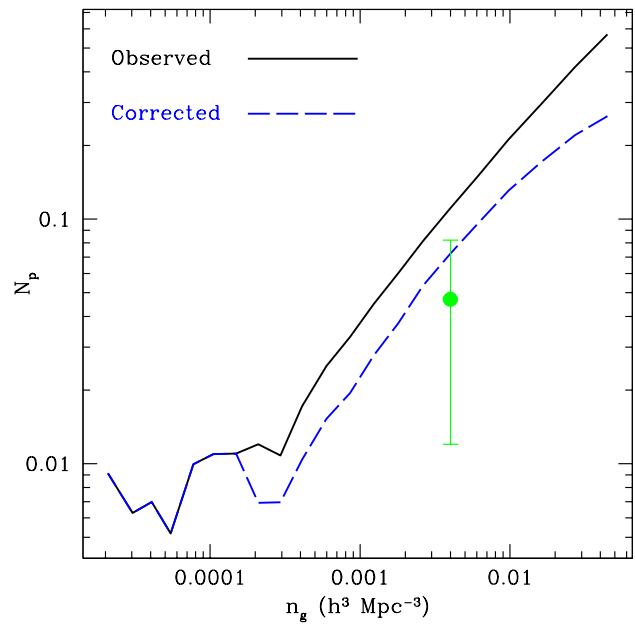}}}
\caption{ The photometric  close pair fraction at $z=3$  as a function
  of the  number density  of galaxies in  the sample. The  solid black
  line  is the  observed photometric  close pair  fraction.   The blue
  dashed line is the close pair fraction for the physically associated
  sample  only. This line  illustrates only  the fraction  of galaxies
  that would  be mutually  inside the same  hosting dark  matter halo.
  The  green  octagonal  point  represents  the  observed  close  pair
  fraction  for  a  sample   of  photometrically  observed  LBGs  from
  \citet{Cooke09} (see text for further details).  }
\label{fig:Z3Phot}
\end{center}
\end{figure}

We   examine  this   solely  at   $z=3$  in   a  fashion   similar  to
Figure~\ref{fig:z3n} for our photometric  pairs sample.  The result is
Figure~\ref{fig:Z3Phot}, which presents $N_p$  as a function of $n_g$.
The solid black  line is the observed $N_p$, and  the blue dashed line
is the fraction of galaxies that are actually physically associated.

%---------------------------------------------------------------------
%
%
%
%
%
%---------------------------------------------------------------------

We find a photometric close pair fraction, $N_p$, of $0.336 \pm 0.009$
before correcting  for impurity.  After  this correction we  find $N_p
\sim 0.189 $, which agrees with our  predicted $N_c$.  The discrepancy
between the ``observed'' and  the physically associated  fraction, the
``pure'' portion of the sample  without line of sight contaminants, is
reasonable due to  the large effective length of  the cylinder used in
these  measurements.   The corrected   fraction reflects the  ``real''
close  ($<30 ~\hkpc$) companion  count and therefore the real fraction
of interacting galaxies.

These results are interesting when compared with the results presented
in  \citet{Conselice03}, \citet{Bertone09}, and  \citet{Bluck09} which
produce similar statistics for these high redshift objects.  The close
pair  fractions  we find  are  consistent  with  the merger  fractions
estimated in these works for  other observed large samples of galaxies
using  close  galaxy   pairs  as  well  as  estimates   based  on  the
concentration-asymmetry-clumpiness  (CAS) method.  \citet{Conselice03}
estimates an apparent merger fractions  of bright LBG's at $z \ge 2.5$
to  be  between  $40-50$  per cent.  \citet{Bluck09}  identifies  $82$
massive galaxies, $M_*  > 10^{11.0} M_{\odot}$ in a  redshift range of
$1.7-3.0$.  These  are further  divided into a  sample of  44 galaxies
between $1.7 < z < 2.3$ and $38$ galaxies from $2.3 < z < 3.0$.  Close
pair fractions  are estimated for  the two samples by  identifying all
imaged  galaxies within $\pm  1.5$ magnitudes  of the  sample galaxies
magnitude that reside within an $R_{max} < 30$ kpc (physical, $h=0.7$)
and statistically corrected for impurities.  Please note that only the
original  $82$ galaxies  have redshift  information, the  rest  of the
galaxies  included  in the  measurement  are  photometric pairs.   The
observed $N_p$ for  the $1.7 < z <  2.3$ ($2.3 < z < 3.0$)  are $N_p =
0.19 \pm 0.07$ ($N_p =  0.40 \pm 0.10$).  These observed high redshift
values are  consistent with  our estimates in  both the  corrected and
uncorrected  sample.   \citet{Bertone09}  examines  models  of  galaxy
merger  rates comparing simulations  and observational  results.  This
work also finds a high merger rate at high redshift.

%----------------------------------------------------------
\subsection{Comparison To Previous Results}
\label{sub:compare}%Section 3.3
%----------------------------------------------------------

In   order  to demonstrate the    usefulness   of  this technique   in
calculating   $N_c$  across a  wide  range  of   redshifts  we  make a
comparison  to several previous existing  measurements at low redshift
(see  \citet{Berrieretal06} for further  discussion on these samples).
The samples used  in  this comparison are  extracted  from the  SSRS2,
CNOC2,  and  the   DEEP2 surveys \cite{P2002,Lin04}.    These  surveys
provide several candidate definitions for a close  pair.  To match the
technique used in  the observations we use criteria  (A).  In order to
determine the sample of haloes used  in our simulation, both host dark
matter   haloes  and  substructure, we  match   the  number density of
galaxies from the observations to the number  density of haloes in the
simulation using the $V_{in}$ model as in \citet{Berrieretal06}.
 
To extend this method to $z\sim3$, we examine observed colour-selected
LBG pairs with $10-30 h^{-1}$ kpc separations from our LBG survey and
the larger survey of S03 and find an observed photometric pair
fraction of $N_p = 0.047 \pm0.035$.  This result is corrected for
spatial impurities that are estimated in a manner similar to that for
the low-z samples.  We estimate the impurity using mock catalogs
constructed to the exact field dimensions and number densities of each
observed field.  We then distribute mock galaxies using redshift
distributions and interloper fractions determined by the photometric
selection function.  The number of close pairs observed in projection
is then corrected to align with the fraction that is found to consist
of true pairs in three dimensions.

We then calculate the $z=3$ data  in the simulation in the same manner
as the  low-redshift data  and matched it  to the LBG  number density.
From our definition  of the close pair fraction,  a decrease in number
density corresponds to a similar  decrease in the close pair fraction.
If LBGs  randomly comprise $\sim1/5$  the number of massive  haloes as
the  number densities  imply,  and thus  approximately  $1/25$ of  our
sample would  be composed  of LBG-LBG pairs,  we would expect  a lower
limit of  $N_c \sim  0.228 \times (1/5)$,  or $\sim0.0456$  ($N_c \sim
0.177 \times (1/5)$, or  $\sim0.0373$ for the purity corrected sample)
for the observed LBG-LBG pair  fraction.  Our simulated value is still
larger than  this at $N_c  = 0.083 \pm  0.005$ ($N_c \sim  0.065$ with
purity  corrections).   For  comparison,  we  look at  $N_p$  for  our
photometric pair value after corrections  for impurity, as this is the
only  estimate of  the full  $N_c$ available.   This assumes  that all
galaxies in the  fiducial simulation sample which are  within the same
host halo will be observed, irregardless of the $n_g$ issues discussed
in \S~\ref{sub:NG}.

%We then calculate the $z=3$ data  in the simulation in the same manner
%as the  low-redshift data  and matched it  to the LBG  number density.
%From our definition  of the close pair fraction,  a decrease in number
%density corresponds to a similar  decrease in the close pair fraction.
%If LBGs  randomly comprise $\sim1/5$  the number of massive  haloes as
%the  number densities  imply,  a nd thus  approximately  $1/25$ of  our
%sample would  be composed  of LBG-LBG pairs,  we would expect  a lower
%limit of $N_c \sim 0.177 \times (1/5)$, or $\sim3.73$ per cent for the
%observed LBG-LBG  pair fraction.  Our simulated value  is still larger
%than this  at $N_c = 0.0506  \pm 0.0017$.  For comparison,  we look at
%$N_p$ for  our photometric pair value after  corrections for impurity,
%as  this is  the  only estimate  of  the full  $N_c$ available.   This
%assumes that all galaxies in  the fiducial simulation sample which are
%within the same host halo  will be observed, irregardless of the $n_g$
%issues discussed in \S~\ref{sub:NG}.

\begin{figure}
\begin{center}
\scalebox{0.43}[0.43]{\rotatebox{0}{\includegraphics{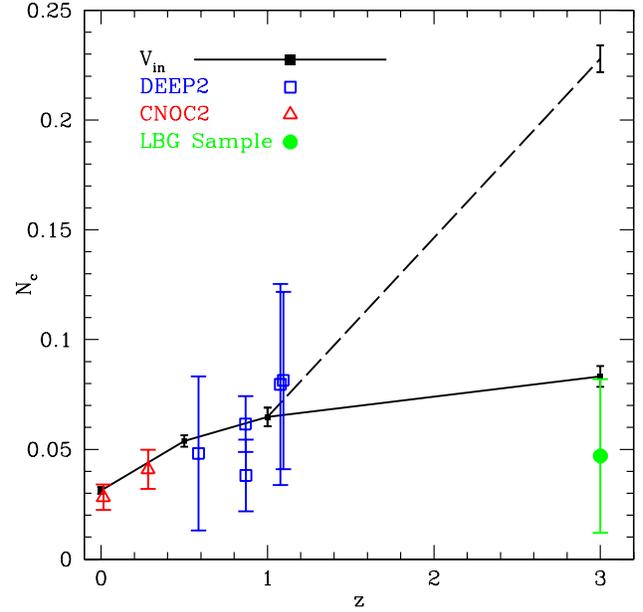}}}
\caption{ Comparison  of $N_c$ from  $z=0$ to $z=3$ in  the simulation
  with measurements from surveys. The black points, solid squares, and
  solid  line represent the  $N_c$ calculated  from our  $n_g$ matched
  samples from $z=0$ to $z=3$.  The dashed black line represents $N_c$
  for our fiducial  sample at $z=3$.  The empty  blue squares are data
  points  from   the  DEEP2  Survey   taken  from  fields  1   and  4,
  \citep{Lin04}.   The hollow red  triangles are  data from  CNOC2 and
  SSRS2, \citep{P2002}.   The green dodecagonal point at  $z=3$ is the
  $N_p$  measured   from  the  LBG  surveys   of  \citet{Cooke05}  and
  \citet{Steidel2003}.  }
\label{fig:compare}
\end{center}
\end{figure}

The    results     of    this    comparison     are    presented    in
Figure~\ref{fig:compare}.   Here the solid  black line,  black points,
and solid  square points represent the pair  fractions calculated from
our $n_g$ matched  samples from $z=0$ to $z\sim3$.   The empty squares
are observed values  from the DEEP2 Survey taken from  fields 1 and 4,
\cite{Lin04} and the  triangles are from the CNOC2  and SSRS2 surveys,
\cite{P2002}.  The  value of $N_c$  from the $z\sim3$  observations is
indicated by the green dodecagonal  point at $z=3$.  The full fiducial
sample $N_c$  for $z=3$  is shown connected  by the dashed  line.  The
$N_c$ values determined from the simulations are closely comparable to
the matched-density  observations in  nearby redshift bins  across all
redshifts  observed.   The  technique  used here  is  more  completely
described in \citet{Berrieretal06}.

%---------------------------------------------------------------------
%
%
%
% 
%
%----------------------------------------------------------------------

%----------------------------------------------------------
\subsection{Galaxy Number Density Problem Revisited}
\label{sub:Density}%Section 3.4
%----------------------------------------------------------

We have matched the 3-D  two-point correlation function to produce our
fiducial  sample of  high redshift  galaxies.   We find  that if  LBGs
comprise  all massive haloes  in the  matched simulation  sample, this
produces a number  density that is too high by  a factor of $\sim4.75$
when  compared with  the  observed  LBG number  density,  as has  been
similarly found by other authors.  In addition, this sample produces a
value  of $N_p$  and $N_c$  in the  cylinders that  is too  high  by a
similar  factor but  a value  of $N_{cs}$  in the  spectroscopic slits
(that are  biased to  detecting pairs with  separations of  $<\sim 20$
h$^{-1}$ kpc because of their  geometry) that is consistent within the
errors of the observations.

In  contrast, forcing  the sample  to  match the  observed LBG  number
density yields an $N_p$ and cylinder $N_c$ similar to the observations
but an $N_{cs}$ that is  significantly lower than that observed in the
slits.  Moreover, such a sample ($V_{in} \gtrsim 200$ \kms) results in
a poor fit to the observed LBG correlation function and corresponds to
haloes  too  massive to  be  reconciled with  the  mass  of LBGs  from
clustering analysis.

LBGs  do  not  comprise  all  massive  galaxies  at  $z\sim3$.   Other
identified  populations  include  sub-mm  galaxies, passive  and  star
forming  BzK galaxies,  DRGs, and  other galaxy  types  with typically
lower UV  luminosities then LBGs.   If LBGs represent  $\sim1/4.75$ of
the matched massive haloes in  our $z\sim3$ sample, then the agreement
in number  density produces  an $N_p$ and  cylinder $N_c$ that  are in
very good agreement with the observations but an $N_{cs}$ in the slits
that is significantly lower than  the observations.  In this case, the
sample still matches the LBG  correlation function (as they are pulled
from the same  parent population) and thus corresponds  to haloes with
the same mass as the observations.

Dust obscured galaxies may make up a fraction of the "missing" massive
haloes,  however, infrared  and sub-mm  surveys recover  only  a small
fraction  of the  number necessary  to reconcile  the  difference. The
latter  interpretation  more  accurately  reflects  the  observed  LBG
population and leaves only the  $\lesssim 20 h^{-1}$ kpc physical pair
fraction  in disagreement.  The  small-scale behaviour  (one-halo term
regime) measured  in accurate  2-D correlation functions  that utilise
deep, wide-field  imaging helps provide the solution  to the remaining
disagreement  and  offers  interesting  insight into  the  nature  and
detectability of LBGs.

%*********************************************************************

%----------------------------------------------------------
\subsection{Small-scale behaviour: The one-halo term}
\label{sub:One-halo}%Section 3.5
%----------------------------------------------------------
 
Measurements  of the   angular  (2-D)  correlation function   (ACF) at
$z\sim4$  in   the Subaru/XMM-Newton Deep  Field   over $1$ deg$^{-2}$
\citep{Ouchi05} and at $z\sim3$  in the Canada-France-Hawaii Telescope
Legacy Survey $4$  square-degree Deep fields  \citep{Cooke12} are able
to utilise a large number ($\sim$$10^4  - 10^5$) of LBGs to accurately
probe the  ACF down   to $m_R\sim27$ from   relatively small  to large
scales ($\sim 0.5 - 10  ~h^{-1}$ Mpc, comoving).  Both efforts witness
a distinct break in the form of the ACF at small scales from the power
law fit    over  larger scales that   may   provide insight   into the
discrepancy in the  observed and expected  close pair fractions in the
spectroscopic slits.

In order to  measure the angular correlation function  at $z=3$ in the
simulation, we will  follow the method used in  previous works such as
\citet{Conroy2006} and utilize the Limber transformation.
\begin{equation}
\omega(\theta)=\frac{\int_0^{\infty}dz N^2(z) \frac{dr}{dz} \int_{\infty}^{\infty}dx \xi(\sqrt(r^2\theta^2+x^2))}{[\int_0^{\infty}dz N(z)]^2},
\label{eqt:limber}
\end{equation}
where $r$ is the comoving distance at $z$ and $N(z)$ is the normalized
redshift distribution of the galaxies in the observed sample.

\begin{figure}
\begin{center}
\scalebox{0.43}[0.43]{\rotatebox{0}{\includegraphics{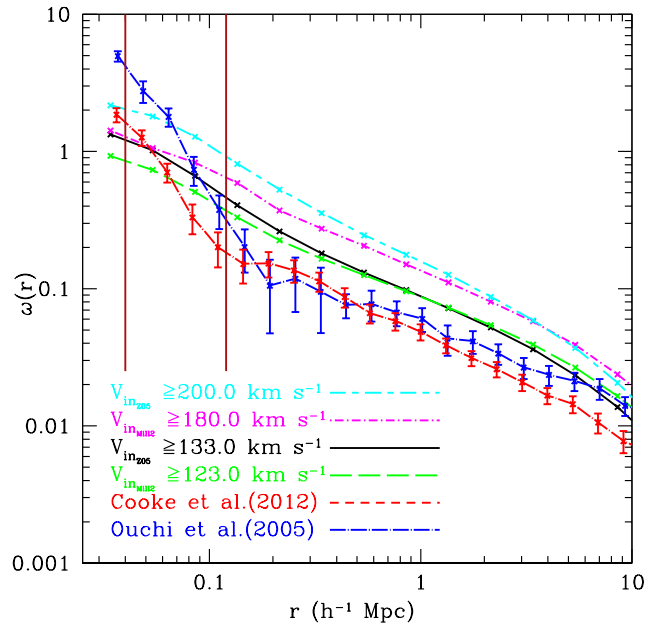}}}
\caption{The angular  correlation function for samples  of dark matter
  haloes  pulled from our  primary sample.   Two subsamples  are taken
  from the dark matter halo  catalog using different $V_{in}$ cuts and
  are compared with  two observational sets.  The solid  black line is
  our fiducial sample which  matches the three dimensional correlation
  function of \citet{Adelberger05}.  The green long-dashed line is our
  best  match to  data from  the Millennium-2  simulation  sample with
  $V_{in} \ge 123$ km s$^{-1}$.  The magenta dot-short dashed line and
  cyan short-long dashed line represent abundance matched samples with
  $V_{in} \ge 180$ km s$^{-1}$  from Millennium-2 and $V_{in} \ge 200$
  km s$^{-1}$  from the Z05  simulations respectively.  The  red short
  dashed sample  represents the observed  angular correlation function
  for $z  \sim 3$ from  \citet{Cooke12}, and the blue  dot-long dashed
  line  illustrates a sample  of LBGs  at $z=4$  from \citet{Ouchi05}.
  Both observational samples demonstrate  a strong break from a single
  power law around $100$ h$^{-1}$ kpc.  None of the simulation samples
  are able  to reproduce this  feature through our  standard abundance
  matching techniques. The two dark red solid vertical lines represent
  the range,  in comoving coordinates, that  we identify galaxy-galaxy
  pairs. }
\label{fig:ang}
\end{center}
\end{figure}

Figure~\ref{fig:ang} presents  the ACFs of our  simulation samples and
the    two    observational     datasets.     As    demonstrated    in
Figure~\ref{fig:ang}, the  form of the fiducial ($V_{in}  > 133$ \kms)
and observational ACFs in  the outer regions are consistent.  However,
the inner regions  show a marked discrepancy.  {\it  Regardless of the
  number  density, the  standard technique  of halo  assignment cannot
  reproduce the  features of the  LBG angular correlation  function on
  both large and  small scales}.  In addition, the  mismatch cannot be
corrected by any scaling of  the data via an integral constraint.  Our
standard abundance matching model  is unable to reproduce the observed
break from power law behaviour near $150-200$ h$^{-1}$ kpc in comoving
coordinates to match both the  one-halo and two-halo components of the
correlation function to the accuracy of the data without incorporating
assumptions of the form of the Limber equation in the inner region.

%----------------------------------------------------------------------
%
%
%
%
%
%---------------------------------------------------------------------

If we follow a standard subhalo abundance matching scheme we observe a
$N_c$ in the cylinder and $N_p$ from our mock photometric sample which
are  consistent with  the observed  pair  counts, however,  we do  not
produce the  correct $N_{cs}$ for the mock  spectroscopic slits (pairs
at  $\lesssim 80  ~\hkpc$  in comoving  coordinates,  or $\lesssim  20
~\hkpc$ in physical coordinates).   $N_c$ is smaller than the observed
fraction by a factor consistent  with the decrease in amplitude of the
one-halo term in the correlation function over the same separations.

To further  test this result,  we analyse the Millennium  2 simulation
\citep{MBK2009}.   The  Millennium-2  simulation  has five  times  the
spatial resolution of the original Millennium simulation, in this case
a Plummer equivalent softening of  $1 ~\hkpc$. This resolution is more
than adequate for  our needs.  The sample we  select from Millennium-2
is  well above  the  resolution  limits of  the  simulation, and  thus
provide a further test that  our results are not adversely effected by
resolution issues. Using this simulation, we find the closest match to
the two point correlation function of  the observations to be a cut of
$V_{in} \ge 123$ km s$^{-1}$.   Haloes with this criterion show a good
agreement to the data,  with a power law fit of $r_0  = 3.98 \pm 1.44$
Mpc h$^{-1}$ and $\gamma =  -1.47 \pm 0.27$.  In addition, this sample
has a  number density $n_g = 0.017$  h$^{3}$ Mpc$^{-3}$, approximately
$4.25$ times larger than the observed $n_g$ of LBGs and similar to the
overdensity of  our primary simulation  sample.  We have  included the
Millennium-2 ACFs  for the  abundance matched sample  and for  the 3-D
correlation function  matched sample in  Figure~\ref{fig:ang}.  Again,
the  simple   abundance  matching   techniques  are  not   capable  of
reproducing the shape of the angular correlation function.

This  selected sample produces  a $N_c  = 0.1521  \pm 0.0006$  for our
fiducial criteria (A) in the  cylinder, and $N_{cs} = 0.033 \pm 0.001$
for  our fiducial  criteria (C)  in the  slits.  The  sample  shows an
impurity  of $0.1433  \pm  0.0176$  in the  cylinder  and $0.1167  \pm
0.0202$  for the  slit,  thus  our corrected  values  are $0.1303  \pm
0.0059$  and  $0.02919 \pm  0.0014$  for  the  full cylinder  and  the
spectroscopic  slit   respectively.   For  our  line   of  sight  mock
photometric  sample  we  find   $N_p  =  0.2299  \pm  0.0144$,  before
corrections for impurity and $0.1370 \pm 0.0154$ after.

To match the number density of  observed LBGs we would select a sample
of haloes  with $V_{in}  \ge 180$ km  s$^{-1}$.  As with  the previous
number density matched sample this has  a power law fit of $r_0 = 5.17
\pm 2.24$ and $\gamma = -1.56 \pm 0.30$.  This sample produces a close
companion  count  of  only  $N_{cs}   =  0.0101  \pm  0.0023$  in  the
spectroscopic  slit,  $N_{cs}  =  0.0089 \pm  0.0022$  after  impurity
corrections, $N_c  = 0.0505  \pm 0.0060$ for  the full  cylinder, only
$0.0442 \pm  0.0061$ after  being corrected for  impurity, and  $N_p =
0.0682  \pm   0.0092$  photometrically,  $0.0471   \pm  0.0075$  after
correcting for sample impurity.

All  results for our  primary samples  in the  Z05 simulations  and in
Millennium-2    from    here    and    from    \S~\ref{sub:evolution},
\S~\ref{sub:pred} are summarised in Table~\ref{table1}

\begin{table*}
\begin{center}
\caption{$N_c$ and $N_p$ for all samples}
\begin{tabular}{|l|c|c|c|c|c|}
\hline 
Simulation \& Criteria     &     $N_c$       &Corrected $N_c$    &    $N_{cs}$      & Corrected $N_{cs}$ &$ n_g$ \\
\hline
Z05 (A)                    & $0.228\pm0.006$ &  $0.177\pm0.005$  &$0.0293\pm0.0013$& $0.0233\pm0.0011$ & $0.019$ \\ 
Z05 (B)                    & $0.246\pm0.006$ &  $0.193\pm0.005$  &$0.0375\pm0.0016$& $0.0305\pm0.0014$ & $0.019$ \\ 
Z05 (C)                    & $0.275\pm0.007$ &  $0.218\pm0.006$  &$0.0506\pm0.0017$& $0.0419\pm0.0015$ & $0.019$ \\ 
Z05(R)(A)                  &       N/A       &      N/A          &$0.0220\pm0.0234$&        N/A        & $0.019$ \\ 
Z05(R)(B)                  &       N/A       &      N/A          &$0.0300\pm0.0178$&        N/A        & $0.019$ \\ 
Z05(R)(C)                  &       N/A       &      N/A          &$0.0430\pm0.0145$&        N/A        & $0.019$ \\ 
Z05(Photometric)           & $0.336\pm0.009$ &  $0.189\pm0.003$  &       N/A       &        N/A        & $0.019$ \\ 
Millennium-2 (A/C)         &$0.1521\pm0.0006$&$0.1303 \pm 0.0059$& $0.033\pm0.001$ & $0.02919\pm0.0014$& $0.017$ \\ 
Millennium-2 (Photometric) &$0.2299\pm0.0144$&$0.1370 \pm 0.0154$&      N/A        &       N/A         & $0.017$ \\
Observations Photometric   & $0.047\pm0.035$ &      N/A          &      N/A        &       N/A         & $0.004$ \\ 
Observations Spectroscopic &       N/A       &      N/A          & $0.047\pm0.015$ &       N/A         & $0.004$ \\ 
\hline
\end{tabular}
\label{table1}
\end{center}
{\small  The Z05  models are  our standard  simulations.  Millennium-2
  simulation values  are reported using  criteria (A) in  the cylinder
  and criteria  (C) in the slit.  The number density  of the abundance
  matched sample  used in  Figures 8  and 11 are  $n_g =  0.004$ h$^3$
  Mpc$^{-3}$}
\end{table*}

%----------------------------------------------------------
\subsection{Discussion}
\label{sub:Discussion}%Section 3.6
%----------------------------------------------------------

We find that matching haloes in our simulation to the observed LBG
number density or the LBG 3-D correlation function and mass using a
simple prescription can generate informative close pair statistics.
We find that the low density simulation samples are able to reproduce
the total observed $N_c$ and $N_P$ in cylinders, but underpredict the
fraction observed serendipitously in spectroscopic slits.  In
contrast, the higher density 3-D correlation function matched sample
is able to reproduce the spectroscopic slit fraction, but overpredicts
the number of observed galaxy-galaxy pairs.  Our numerical/analytical
simulation is not resolution limited in this sense and the discrepancy
at small scales occurs above the resolution limit of the Millennium-2
simulation.  At these separations, many of the luminous galaxies
sharing these haloes are either interactions or imminent interactions.
This finding provides an interesting avenue to quantify the spatial
behaviour of LBGs and sub-halo assignment schemes.

Galaxy interactions  can generate  a significant enhancement  in their
luminosities         from        the         close        interactions
\cite[e.g.][]{LarsonTinsley78,    Barton:00,    Ellison08,   Bridge10,
  Wong11}.  Moreover,  it is possible that  the luminosity enhancement
at high  redshift is equivalent to,  or higher than,  that observed at
low redshift as a result of the higher gas fractions in LBGs.

The  Ly$\alpha$ emission  versus separation  relationship  of observed
close  LBG pairs  found  in C10  supports  this picture.   All of  the
spectroscopic  $\lesssim20  ~\hkpc$  physical  close pairs,  and  thus
interacting  systems,  exhibit  Ly$\alpha$  emission  as  compared  to
$\sim50$ per cent of the  full population.  A fraction of the observed
Lya emission of each interacting galaxy is likely to be a signature of
enhanced star formation.  This  behaviour may extend to the Ly$\alpha$
emitter (LAE) population as well (see C10).

Lower-luminosity     LBGs      typically     have     lower     masses
\citep{Giavalisco2001,  Kashikawaetal2006} and  the higher  density of
these haloes yields  a higher interaction fraction as  compared to our
fiducial  (m$_R<25.5$)   sample.   The  luminosity   enhancement  from
interactions would boost a fraction of lower-luminosity LBGs above the
magnitude selection cut-off of  our sample.  This process would create
an increase in the number  of $\lesssim20 ~\hkpc$ physical close pairs
detected  in  the  observations   that  are  not  represented  in  the
simulation analysis.

Our adopted halo  assignment (\S 2.3),  does not account for  enhanced
star  formation.  Moreover our  $V_{now}$  sample (\S~\ref{sec:model})
models haloes where baryons are stripped during infall.  Such galaxies
would have a decrease in luminosity as compared to our $V_{in}$ model,
and we find the $V_{now}$ model predicts fewer close pairs detected in
the slits.  This result implies that a model which instead includes an
appropriate luminosity enhancement  per  baryon for infalling   haloes
over   the standard assumptions of  abundance  matching will predict a
higher  fraction  of close  pairs in  the   slits  as is  seen  in the
observations.

Our results and  the proposed scenario  remain consistent within  high
redshift measurements and   fractions of close  or interacting/merging
LBGs  by    various means when      considering  the samples   studied
\citep[e.g.,][]{Conselice03, Lotz06a, Bertone09,  Forster-Schreiber09,
Bluck09,  Law12}.  In   addition,  observations  of low  redshift  LBG
analogs \citep{Overzier09,Overzier10,Goncalves10}, which  are  matched
to LBGs  in essentially every   way (stellar mass, gas  fraction, star
formation   rate,  metallicity, dust  extinction,  physical  size, gas
velocity dispersion, etc.), show from optical imaging that the bulk of
these objects are undergoing  interactions even though the UV  imaging
is inconclusive.   Simulated  to  high  redshift, these   objects  are
consistent with the properties and observations of $z\sim3$ LBGs.

\citet{Law12}  uses Hubble  Space Telescope  imaging in  the restframe
optical to  estimate the  number of  real close pairs  in a  sample of
galaxies  observed  between  $2.5  \le  z  \le  3.6$.   Galaxies  with
spectroscopically   determined   redshifts   and  magnitudes   between
$H=22.0-24.0$ were compared to objects within $5-16$ h$^{-1}$ kpc with
no   more  then   one  magnitude   difference.   These   results  were
statistically corrected for  false close pairs and produce  a value of
$N_c  =  0.17^{+0.12}_{-0.08}$ for  $z\sim3$  LBGs.   This close  pair
fraction is consistent with our results.

In our fiducial model, selected  by matching the two point correlation
function, we have not truly required all galaxies to be visible either
due to dust extinction or  low star formation, thus  we do not have to
match the observed number density  as we would  in Sub Halo  Abundance
Matching.  The work of  \citet{Reddy2008} suggests that the rest frame
UV  luminosity of galaxies at  these redshifts are typically extincted
by  a factor of $4-5$  in flux.  Our  typical halo masses at accretion
are $M \ge 10^{11.54}$. All but $\sim 1$ per cent of  our halos have a
mass at accretion above the minimum  mass of $M_{min} \ge 10^{11.1 \pm
0.2}$ h$^{-1}$ M$_{\odot}$  suggested  by halo modeling  for  $z\sim2$
star forming galaxies  in \citet{Conroy2008}, where the number density
of halos does not match the observed $n_g$ for that population.

Our number-density matched  sample  implies that every halo  hosts  a
luminous  LBG.  In  this case,  the lower  fraction of close  pairs at
$\lesssim20$ kpc  separations   as compared to  the  observations goes
against interacting galaxy  behaviour.   Moreover, we  know  from high
redshift surveys that   LBGs comprise a large  fraction,  but not all,
detectable galaxies  at  high redshift   and that  haloes indeed  host
massive galaxies that are not detectable using Lyman break techniques.
Thus, we are forced to consider higher densities samples.  In order to
generate  4.75 times  the  observed LBG  density  of  the  correlation
function matched sample, we  need to integrate  down the  faint-end of
the luminosity   function   below   $R=25.5$  by  $\sim$1    magnitude
\citep{sawicki06,Reddy2008}.   Although   a   fraction of the   higher
luminosity haloes   in   this magnitude  range  are  expected  to have
sufficient star formation enhancement to enter into our magnitude cut,
the  lack of  knowledge  of  the  typical  enhancement in  the  FUV at
$z\sim3$ makes it unclear if such a  fraction is large enough to match
that  observed in the  spectroscopic slits.  If a significant fraction
of the  star  forming galaxies are   obscured by dust  as suggested by
\citet{Lacey2011}, a random   subsample would  have approximately  the
same correlation  function and produce   similar close pair fractions.
As  a result,  the enhanced  fraction of  lower-luminosity galaxies may
include high star formation rate, dusty galaxies that may experience a
larger   magnitude  increase from the      effects of interaction  and
morphological disruption.

Finally, assuming that LBGs comprise $\gtrsim$50\% of all star forming
galaxies at $z\sim3$  \citep{Reddy05,Marchesini2007},  the  results of
the correlation function matched sample suggest that after considering
dust-obscured galaxies, $\sim$3/5 of haloes are not observed using any
high  redshift detection technique.   Driven  by (1)  the power of the
simulation   to  predict the  close     pair fraction  down  to  $z=0$
\citep{Berrieretal06}, (2) the evidence that abundance matching may be
used with  high   redshift LBGs \citep{Conroy2006}, (3)  the  possible
direct correlation  between halo mass and UV  luminosity at this epoch
\citep{Simha2010,Conroy2009}, and  (4)  the equivalent  overdensity of
similarly  matched  massive haloes   in  other simulations   including
simulations  using different  approaches \citep[e.g][]{Lacey2011}, the
unaccounted $z=3$ haloes  in  the correlation function matched  sample
likely  reflect a similar  number of real  haloes in the Universe.  If
true, these haloes must   either have highly obscured  star  formation
that is not detected  by  current high redshift selection   techniques
\citep[e.g.][]{Lacey2011} such as IR and sub-mm  surveys, or they must
be massive galaxies with inherently low  star formation rates that are
below  the detection thresholds of  current facilities.  It may be the
case that galaxy interactions is  the cause for the initial starburst,
or ``turn-on", of many of  these undetected haloes within our  sample.
Thus, a  combination  of  all of   the above  affects   resulting from
interaction induced star formation may provide a plausible explanation
for the larger fraction of observed ($<20$ kpc) serendipitous pairs in
spectroscopic slits  when assuming  LBGs  comprise $\sim$1/5   of  the
correlation  function  and  mass  matched  sample  and   $\sim$1/5 the
reported close pair fractions.  One means to probe such massive haloes
independently of   their  luminosity is  via   quasar absorption  line
systems,    in particular,  the     ubiquitous  damped  \lya\  systems
\citep{Wolfe2005}, which have been shown to be associated with massive
systems     that          cluster       similar       to          LBGs
\citep{Cooke06a,Cooke06,Nagamine06,Lee08}.   We  are    engaged     in
investigation that  is testing various components  of this scenario on
several fronts.

%*********************************************************************

%----------------------------------------------------------
\section{Conclusions} 
\label{sec:conclusions}%Section 4
%----------------------------------------------------------

We have matched the 3-D two  point correlation function of a sample of
$z   \sim  3$  LBGs   to  a   sample  of   haloes  from   our  primary
numerical/analytic cosmological simulation.  Using this sample we have
mocked   observations  of   simulated  spectroscopic   slits   and  of
photometric observations  of these galaxies.   We also test  our model
with  data from the  Millennium-2 simulation  for verification  of our
results.  This work has led us to several interesting results which we
summarise in the points below, see also Table \ref{table1} above.

\begin{itemize}

\item We demonstrate that neither standard SubHalo Abundance Matching
  (SHAM) or a two-point correlation function and mass matching scheme
  completely reproduces the observational  results. Neither  model can
  reproduce   galaxy   clustering     features  and   $n_g$   at  same
  time. Explicitly, we find that the standard  SHAM does not reproduce
  the  serendipitously  observed $N_{cs}$, and  the  break  in the LBG
  correlation  function  at  very   small   scales ($\lesssim20~\hkpc$
  physical, $\sim80~\hkpc$ comoving).

\item The number  density of our candidate LBG  sample is $\sim$$4.75$
  times the observed LBG number density.  The implication is that only
  $\frac{1}{5}$ halos  above a fixed mass  are detectable LBGs.  These
  results    are     consistent    with       the     results       of
  \citet{Nagamine2004,Nagamine06,Lee08, Dave2000,Lacey2011} and others
  which find a similar overdensity using other types of simulations.

\item We find an observed close pair fraction $N_c = 0.228 \pm 0.006$,
  which implies  an impurity corrected  close pair fraction of  $N_c =
  0.177 \pm 0.005$, ($\sim18$ per cent) This result is consistent with
  the   previous   results   of   \citet[][]{Conselice03,   Bertone09,
  Bluck09,Law12,    Lotz06,    Forster-Schreiber09}   considering  the
  uncertainties specific  to those  studies and  pair  fraction/merger
  rate assumptions.

\item Our  simulated matched spectroscopic slits produce  a close pair
  fraction of  $N_{cs} = 0.0506  \pm 0.0017$ for our  fiducial sample,
  defined to have a maximum  velocity separation of $\pm 500$ \kms and
  an on the  sky separation of $\le 30.0 ~\hkpc$.   This is similar to
  the observed fraction of  serendipitous spectroscopic close pairs of
  $N_{cs} =  0.047 \pm 0.015$ for  the full observed  LBG sample and
  $N_{cs} =  0.071 \pm  0.023$ for the  highest signal to  noise ratio
  sample. After correcting for false close pairs which may be observed
  we find $N_c = 0.0419 \pm 0.0015$ in the simulation.

\item If we  examine our catalogs using randomly  selected slitlets to
  generate a more  generic result, we find that  the expected fraction
  of  LBG pairs  that fall  serendipitously  into the  slitlets to  be
  $N_{cs} = 0.0430 \pm 0.015$ before correcting for sample impurity.

\item We find  a photometric close pair fraction of  $N_p = 0.336 \pm
  0.009 $  and after correcting for  the sample  impurity we find $N_p
  \sim 0.189  $.  The latter  fraction reflects the ``real'' number of
  close pairs and therefore the real number of potentially interacting
  galaxy pairs.  As  mentioned above, only a portion  of the corrected
  value will be observable LBGs.   The difference in $n_g$ is a factor
  of $\sim 4.75$ leading to a  corrected value of $N_p \sim 3.99 $ per
  cent, which is consistent with our photometric pair fractions $N_p =
  0.047  \pm0.035$   estimated  from  our survey  and   the  survey of
  \citet{Steidel2003}.

\item  The analysis  of the  sample taken  from  Millennium-2 produces
  similar results to our primary simulation.  In Millennium-2, we find
  the   correlation  function-matched  sample   to  be   overdense  in
  comparison with  the observations by  a factor of  $\sim4.25$.  This
  selected  sample  produces  a  $N_c  = 0.1521  \pm  0.0006$  in  the
  cylinder, and $N_{cs}  = 0.033 \pm 0.001$ in  the slits.  The sample
  shows an impurity of $I =  0.1433 \pm 0.0176$ in the cylinder and $I
  = 0.1167  \pm 0.0202$  for the slit,  producing corrected  values of
  $N_c = 0.1303 \pm 0.0059$ and  $N_{cs} = 0.02919 \pm 0.0014$ for the
  full cylinder and the spectroscopic slit respectively.  For our line
  of sight mock photometric sample  we find $N_p = 0.2299 \pm 0.0144$,
  without corrections for impurity and $N_p = 0.1370 \pm 0.0154$ after
  correction.

\end{itemize}

The  excess  of   close (interacting)   pairs $\le    20$ h$^{-1}$ kpc
(physical) and the inability for the  standard abundance matching with
monotonic UV-mass  halo assignment to describe the  steep slope in the
observed  LBG  correlation  function  at very   small  scales provides
insight into triggered star   formation and the detectability  of LBGs
(and  LAEs) at $z=3$.  Our results  imply that  the spectroscopic slit
close   pair fraction  and the  break    in the correlation   function
represent the detection  of either a fraction  of less massive (higher
density) LBGs with luminosities below  our magnitude cut  ($m_R=25.5$)
as  a result of  an  enhancement in  luminosity from interactions, the
``turn-on" of massive  haloes with  previous low  star formation as  a
result of interaction, or, likely, a combination of both cases.

We find that LBGs likely represent $\sim20-25$ per cent of all massive
($V_{in} > 133$ km sec$^{-1}$) haloes at $z\sim3$ based on the results
of  the analysis  of    our simulation, the   Millennium-2 simulation,
simulation  analyses by  several other  authors, and the  power of our
simulation analysis to predict the  close pair fraction from $z=1$  to
$z=0$.  The full census of  detected star forming galaxies selected by
various criteria suggest that  LBGs likely account for $\gtrsim50$ per
cent of the  massive haloes at  $z\sim3$.   The remaining  fraction is
likely   populated by  systems  with  low star  formation rates and/or
systems that   are  not detected using  current  selection techniques.
DLAs are a  promising  means  to explore   the remaining fraction   of
massive haloes because they probe  galaxy haloes randomly, independent
of luminosity, they have a high number  density, and they are found to
reside   in   massive  haloes  \citep{Cooke06,  Fynbo2003,  Fynbo2008,
Fynbo2010, Fynbo2011, Moller2002, Moller2004, Schaye2001}.

The statistics  generated from our  mock spectroscopic slits  with the
serendipitously  confirmed close  pairs from  observations  provides a
potentially powerful tool to estimate the behaviour and nature of LBGs
and the enhanced star formation rate from LBG interactions.

\section*{Acknowledgments}

The simulation was  run on the Columbia machine  at NASA Ames (Project
PI: Joel  Primack).  We thank  Anatoly Klypin and Brandon  Allgood for
running the simulation  and making it available to  us.  We would also
like to thank  Andrew Zentner for providing us  with the halo catalogs
generated by his analytic model. Berrier is currently supported by the
University  of  Arkansas.   The  authors  would like  to  thank  James
Bullock, Elizabeth  Barton, Mike Boylan-Kolchin,  and Kentaro Nagamine
for  useful discussions.  The  authors were  supported in  part during
this work by the Center for Cosmology at the University of California,
Irvine.   J.  C.   gratefully  acknowledges generous  support by  Gary
McCue.  The Millennium-II simulation  databases used in this paper and
the web  application providing online access to  them were constructed
as  part  of  the  activities  of  the  German  Astrophysical  Virtual
Observatory.  The  authors wish to recognize and  acknowledge the very
significant cultural role  and reverence that the summit  of Mauna Kea
has always had within the  indigenous Hawaiian community. 

\bibliography{ms}

\bsp

\label{lastpage}

\end{document}